\documentclass{IEEEtran}
\usepackage{latexsym}
\usepackage{setspace}
\usepackage{cite}
\usepackage{amssymb}
\usepackage{bbm}
\usepackage{algorithmicx}
\usepackage{algorithm}
\usepackage[noend]{algpseudocode} %
\usepackage{bm}
\usepackage{mathtools}
\usepackage[usenames,dvipsnames]{xcolor}
\usepackage{stmaryrd}
\usepackage{amsmath, amssymb,accents}
\usepackage{dsfont}
\usepackage{comment}
\usepackage[dvips]{graphics}
\DeclareMathOperator{\supp}{supp}
\usepackage{graphicx}
\usepackage{lipsum}
\usepackage{psfrag}
\usepackage{units}
\usepackage{subfig}
\usepackage{setspace}
\usepackage{theorem}
\usepackage{float}
\usepackage{mathtools}
\allowdisplaybreaks

\newtheorem{definition}{Definition}

\newtheorem{theorem}{Theorem}
\newtheorem{lemma}{Lemma}
\newtheorem{remark}{Remark}

\newtheorem{corollary}{Corollary}
\newtheorem{proposition}{Proposition}

\newcommand\numberthis{\addtocounter{equation}{1}\tag{\theequation}}
{} 
\begin{document}

\makeatletter
\algnewcommand{\LineComment}[1]{\Statex\hskip\ALG@thistlm \(\triangleright\) #1}
\makeatother

\makeatletter
\newcommand{\vasti}{\bBigg@{3.5}}
\newcommand{\vast}{\bBigg@{4}}
\newcommand{\Vast}{\bBigg@{5}}
\newcommand{\Vastt}{\bBigg@{7}}
\makeatother
\newcommand{\be}{\begin{equation}}
\newcommand{\ee}{\end{equation}}
\newcommand{\ba}{\begin{align}}
\newcommand{\ea}{\end{align}}
\newcommand{\baa}{\begin{align*}}
\newcommand{\eaa}{\end{align*}}
\newcommand{\bea}{\begin{eqnarray}}
\newcommand{\eea}{\end{eqnarray}}
\newcommand{\beaa}{\begin{eqnarray*}}
\newcommand{\eeaa}{\end{eqnarray*}}
\newcommand{\p}[1]{\left(#1\right)}
\newcommand{\pp}[1]{\left[#1\right]}
\newcommand{\ppp}[1]{\left\{#1\right\}}
\newcommand{\ber}{$\ \mbox{Ber}$}
\newcommand{\mkv}{-\!\!\!\!\!\minuso\!\!\!\!\!-}

\title{Wiretap and Gelfand-Pinsker Channels Analogy and its Applications}

\author{Ziv Goldfeld and Haim H. Permuter

\thanks{
		The work of Z. Goldfeld and H. H. Permuter was supported by the Israel Science Foundation (grant no. 2012/14), the European Research Council under the European Union's Seventh Framework Programme (FP7/2007-2013) / ERC grant agreement n$^\circ$337752, and the Cyber Center at Ben-Gurion University of the Negev. Z. Goldfeld was also supported by the Rothschild postdoc fellowship and by a grant from Skoltech--MIT Joint Next Generation Program (NGP).
		\newline This paper was presented in part at the 2018 International Zurich Seminar (IZS-2018), Zurich, Switzerland, and in part at the 2018 IEEE International Symposium on Information Theory (ISIT-2018), Vail, Colorado, US.
		\newline Z. Goldfeld is with the Department of Electrical Engineering and Computer Science, Massachusetts Institute of Technology, Cambridge, MA 02139 USA (e-mail: zivg@mit.edu). H. H. Permuter are with the Department of Electrical and Computer Engineering, Ben-Gurion University of the Negev, Beer-Sheva, Israel (e-mail: haimp@bgu.ac.il).
		}}
\maketitle


\begin{abstract}
An analogy framework between wiretap channels (WTCs) and state-dependent point-to-point channels with non-causal encoder channel state information (referred to as Gelfand-Pinker channels (GPCs)) is proposed. A good sequence of stealth-wiretap codes is shown to induce a good sequence of codes for a corresponding GPC. Consequently, the framework enables exploiting existing results for GPCs to produce converse proofs for their wiretap analogs. 
The analogy readily extends to multiuser broadcasting scenarios, encompassing broadcast channels (BCs) with deterministic components, degradation ordering between users, and BCs with cooperative receivers. Given a wiretap BC (WTBC) with two receivers and one eavesdropper, an analogous Gelfand-Pinsker BC (GPBC) is constructed by converting the eavesdropper's observation sequence into a state sequence with an appropriate product distribution (induced by the stealth-wiretap code for the WTBC), and non-causally revealing the states to the encoder. The transition matrix of the state-dependent GPBC is extracted from WTBC's transition law, with the eavesdropper's output playing the role of the channel state. Past capacity results for the semi-deterministic (SD) GPBC and the physically-degraded (PD) GPBC with an informed receiver are leveraged to furnish analogy-based converse proofs for the analogous WTBC setups. This characterizes the secrecy-capacity regions of the SD-WTBC and the PD-WTBC, in which the stronger receiver also observes the eavesdropper's channel output. These derivations exemplify how the wiretap-GP analogy enables translating results on one problem into advances in the study of the other.

\end{abstract}




\section{Introduction}\label{SEC:introduction}
 

\begin{figure}[!t]
	\begin{center}
	    \begin{psfrags}
	        \psfragscanon
	        \psfrag{A}[][][1]{\ \ $M$}
	        \psfrag{J}[][][1]{\ \ $M$}
	        \psfrag{B}[][][0.9]{$\mspace{-2mu}\mathsf{Enc}$}
	        \psfrag{K}[][][0.9]{$\mspace{-2mu}\mathsf{Enc}$}
	        \psfrag{C}[][][1]{\ \ \ $\mathbf{X}$}
	        \psfrag{L}[][][1]{\ \ \ $\mathbf{X}$}
	        \psfrag{D}[][][1]{\ \ \ \ \ \ \ $p^n_{Y,Z|X}$}
	        \psfrag{M}[][][1]{\ \ \ \ \ \ \ $q^n_{Y|X,Z}$}
	        \psfrag{E}[][][1]{\ \ \ \ $\mathbf{Y}$}
	        \psfrag{N}[][][1]{\ \ \ \ $\mathbf{Y}$}
	        \psfrag{F}[][][0.9]{$\mathsf{Dec}$}
	        \psfrag{O}[][][0.9]{$\mathsf{Dec}$}
	        \psfrag{G}[][][1]{\ \ $\hat{M}$}
	        \psfrag{P}[][][1]{\ \ $\hat{M}$}
            \psfrag{H}[][][1]{${\color{red!65!black}\mathbf{Z}}$}
            \psfrag{Z}[][][1]{${\color{red!65!black}\mathbf{Z}}$}
	        \psfrag{R}[][][1]{\ \ \ ${\color{red!65!black}q_Z^n}$}
	        \psfrag{Q}[][][1]{${\color{red!65!black}\mathbf{Z}}$}
	        \psfrag{X}[][][1]{\ }
	        \psfrag{Y}[][][1]{\ }
	        \psfrag{U}[][][1]{\ }
	        \psfrag{V}[][][1]{\ }
	        \subfloat[]{\includegraphics[scale = .37]{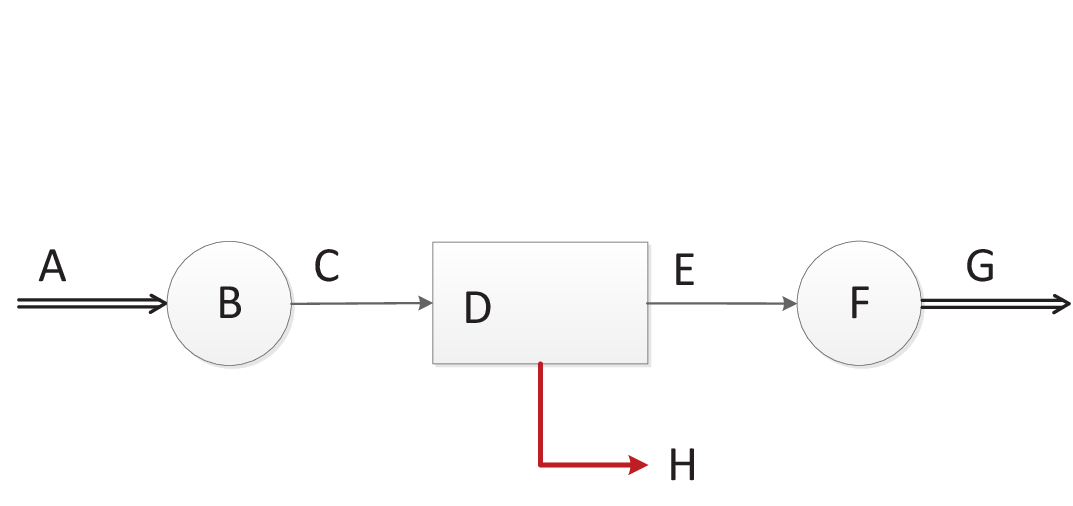}}\hspace{20mm}
	        \subfloat[]{\includegraphics[scale = .37]{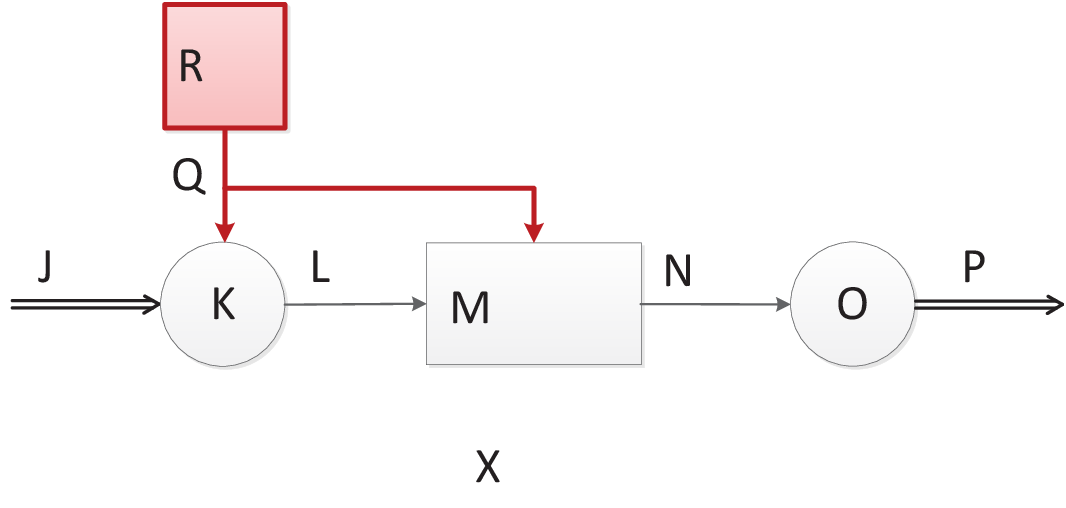}}
	        \caption{(a) The WTC $p_{Y,Z|X}$, where $X$ is the channel input and $Y$ and $Z$ are the channel outputs observed by the legitimate receiver and the eavesdropper, respectively; (b) The GPC with state distribution $Z\sim q_Z$, and channel transition probability $q_{Y|X,Z}$, where $X$ is the input and $Y$ is the output.}\label{FIG:WT-GP_intro}
	        \psfragscanoff
	    \end{psfrags}
	\end{center}\vspace{-7mm}
\end{figure}


Two fundamental, yet seemingly unrelated, information-theoretic models are the wiretap channel (WTC) and the state-dependent point-to-point channel with non-causal encoder channel state information (CSI). The discrete and memoryless (DM) WTC (Fig.~\ref{FIG:WT-GP_intro}(a)) was introduced by Wyner in 1975~\cite{Wyner_Wiretap1975} and initiated the study of physical layer security. Csisz{\'a}r and K{\"o}rner characterized the secrecy-capacity of the WTC \cite{Csiszar_Korner_BCconfidential1978} as
\begin{equation}
    C_\mathsf{WT}(p_{Y,Z|X})=\max_{p_{U,X}}\Big[I(U;Y)-I(U;Z)\Big],\label{EQ:WTC_capacity_intro}
\end{equation}
where $p_{Y,Z|X}$ is the WTC's transition matrix and the underlying distribution is $p_{U,X}p_{Y,Z|X}$. The state-dependent channel with non-causal encoder CSI is due to Gelfand and Pinsker (GP)~\cite{Gelfand_Pinsker}; we henceforth refer to it as the GP channel (GPC) and it is shown in Fig.~\ref{FIG:WT-GP_intro}(b). The capacity of a GPC $q_{Y|X,Z}$ with state distribution $q_Z$ is:
\begin{equation}
    C_\mathsf{GP}(q_Z,q_{Y|X,Z})=\max_{q_{U,X|Z}}\Big[I(U;Y)-I(U;Z)\Big],\label{EQ:GPC_capacity_intro}
\end{equation}
where the joint distribution is $q_Zq_{U,X|Z}q_{Y|X,Z}$. An interesting question is whether the resemblance of \eqref{EQ:WTC_capacity_intro} and \eqref{EQ:GPC_capacity_intro} is coincidental or is there an inherent connection between these problems.

\subsection{The Analogy and its Implications}

This paper shows that an inherent connection is indeed the case by establishing a rigorous uni-directional relation between the WTC and the GPC. 
Specifically, we prove that any good sequence of stealth-wiretap codes for the WTC induces a good sequence of codes of the same rate for a corresponding GPC. This GPC depends both on the original WTC and on the deployed stealth-wiretap code. We add the stealth condition \cite{Kramer_EffectiveSecrecy2014,endo2014reliability} to the regular security requirement of a vanishing information leakage.\footnote{To the best of our knowledge, in all WTC models for which the secrecy-capacity is known both without (e.g., weak- or strong-secrecy) and with a stealth condition, the two capacity expressions coincide.} This allows relating WTCs to GPCs with i.i.d. state sequences (although our  analogy framework is also valid for arbitrary GP state distributions, in which case stealth is not needed). The importance of the i.i.d. state sequence scenario in the GP model stems from the fact that all known single-letter GP capacity results are under this assumption; a partial list includes \cite{GePi80,Lapidoth_senideterministic2012,Steinberg_Degraded_BC_State2005,Dikstein_PDBC_Cooperation2016}. The established relation enables exploiting such known upper bounds on the GP capacity to upper bound the secrecy-capacity of the analogous wiretap model (when achievability is defined to include the stealth requirement). While the solutions to the base cases from Fig.~\ref{FIG:WT-GP_intro} have been known for decades, many multiuser extensions of these models remain open problems. Through the analogy, we derive converse proofs for several multiuser wiretap settings which lead to new secrecy-capacity results.

\begin{figure}[!t]
	\begin{center}
	    \begin{psfrags}
	        \psfragscanon
	        \psfrag{A}[][][0.9]{$(M_1,M_2)$}
	        \psfrag{J}[][][0.9]{$(M_1,M_2)$}
	        \psfrag{B}[][][1]{$\mspace{-2mu}\mathsf{Enc}$}
	        \psfrag{K}[][][1]{$\mspace{-2mu}\mathsf{Enc}$}
	        \psfrag{C}[][][1]{\ \ \ $\mathbf{X}$}
	        \psfrag{L}[][][1]{\ \ \ $\mathbf{X}$}
	        \psfrag{D}[][][0.9]{\ \ \ \ \ \ \ \ $p^n_{Y_1,Y_2,Z|X}$}
	        \psfrag{M}[][][0.9]{\ \ \ \ \ \ \ \ $p^n_{Y_1,Y_2|X,Z}$}
	        \psfrag{E}[][][1]{\ \ \ \ $\mathbf{Y}_1$}
	        \psfrag{N}[][][1]{\ \ \ \ $\mathbf{Y}_1$}
	        \psfrag{U}[][][1]{\ \ \ \ $\mathbf{Y}_2$}
	        \psfrag{X}[][][1]{\ \ \ \ $\mathbf{Y}_2$}
	        \psfrag{F}[][][0.9]{$\mathsf{Dec}1$}
	        \psfrag{O}[][][0.9]{$\mathsf{Dec}1$}
            \psfrag{V}[][][0.9]{$\mathsf{Dec}2$}
	        \psfrag{Y}[][][0.9]{$\mathsf{Dec}2$}
	        \psfrag{G}[][][1]{\ \ $\hat{M}_1$}
	        \psfrag{P}[][][1]{\ \ $\hat{M}_1$}
	        \psfrag{W}[][][1]{\ \ $\hat{M}_2$}
	        \psfrag{Z}[][][1]{\ \ $\hat{M}_2$}
            \psfrag{H}[][][1]{\quad\quad\quad\quad\quad\quad\quad\ \ {\color{red!65!black}$\mathbf{Z}\sim P_{\mathbf{Z}|M_1,M_2}\approx q_Z^n$}}
            \psfrag{R}[][][1]{\ \ \ ${\color{red!65!black}q_Z^n}$}
	        \psfrag{Q}[][][1]{${\color{red!65!black}\mathbf{Z}}$}	        
	        \psfrag{S}[][][1]{\ }
	        \subfloat[]{\includegraphics[scale = .4]{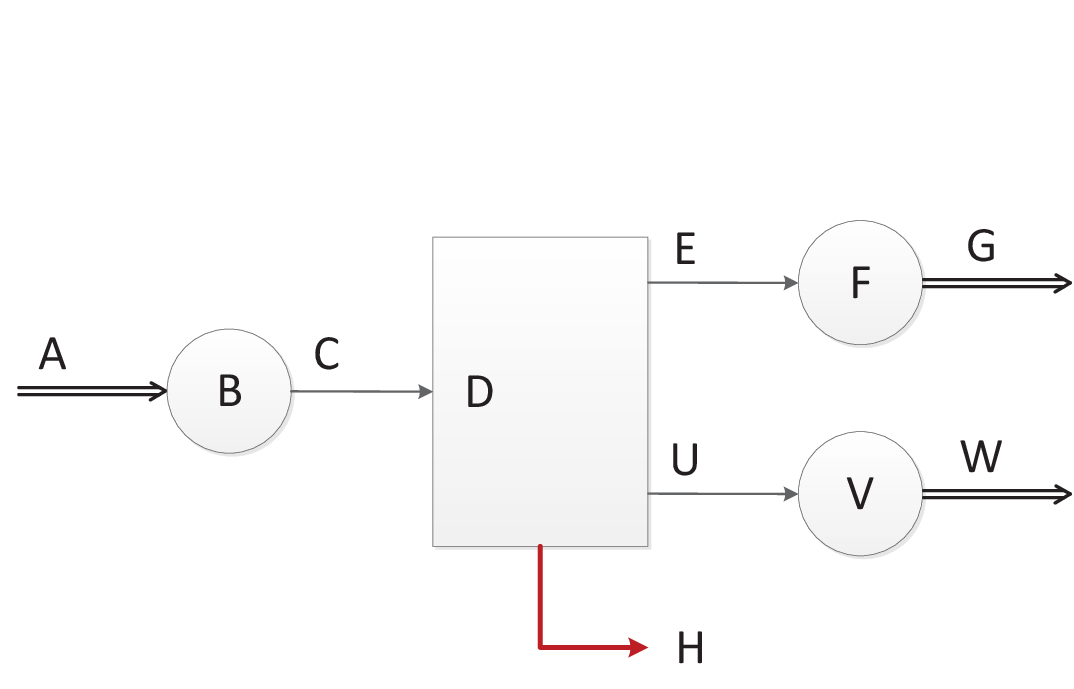}}\hspace{15mm}
	        \subfloat[]{\includegraphics[scale = .4]{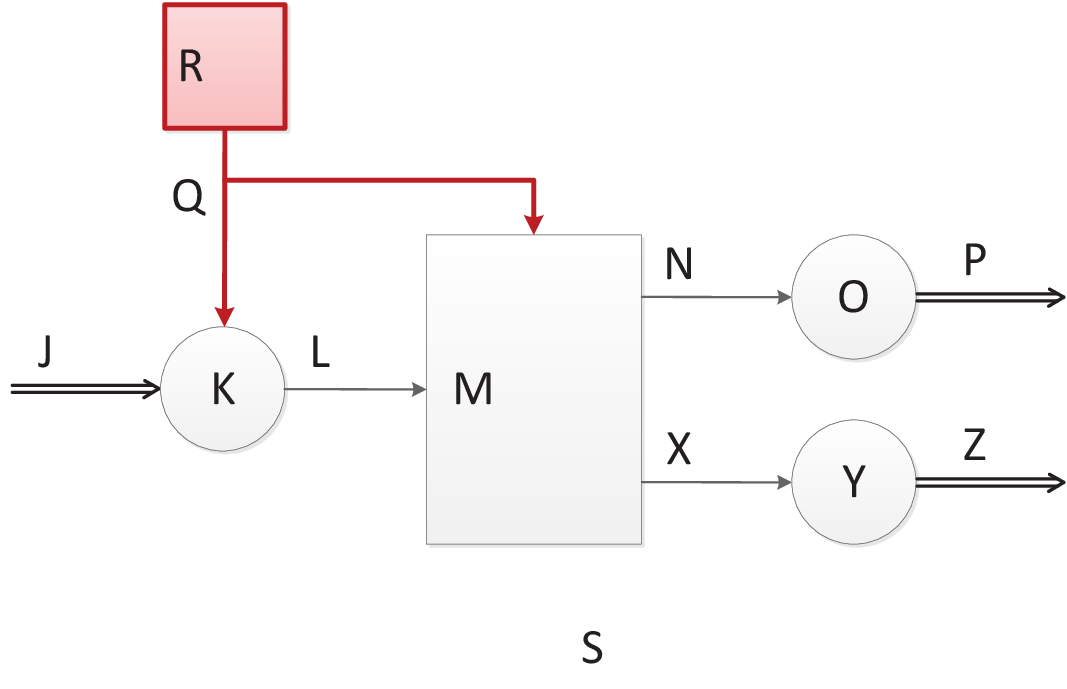}}
	        \caption{(a) The WTBC $p_{Y_1,Y_2,Z|X}$, where $X$ is the channel input and $Y_1$, $Y_2$ and $Z$ are the channel outputs observed by the legitimate receivers and the eavesdropper, respectively. Secrecy and stealth are required versus the eavesdropper, i.e., that $\mathbf{Z}$ is asymptotically independent of the messages and that its distribution, as induced by the code, approximates a target product measure $q_Z^n$. (b) An analogous GPBC is constructed by replacing the eavesdropper's observation with a state sequence $\mathbf{Z}\sim q_Z^n$, non-causally revealing $\mathbf{Z}$ to the encoder and setting the state-dependent BC's transition kernel to $p_{Y_1,Y_2|X,Z}$, i.e., the conditional marginal distribution of the WTBC's law $p_{Y_1,Y_2,Z|X}$.}\label{FIG:WT-GP-BC_intro}
	        \psfragscanoff
	    \end{psfrags}
	\end{center}\vspace{-7mm}
\end{figure}

To this end we extend the wiretap-GP analogy to multiuser broadcasting scenarios. Fix a wiretap broadcast channel (WTBC) $p_{Y_1,Y_2,Z|X}$ (Fig.~\ref{FIG:WT-GP-BC_intro}(a)), with two legitimate receivers observing $Y_1$ and $Y_2$ and one eavesdropper that intercepts $Z$. Our stealth-wiretap codes for the WTBC ensure reliability, security and stealth \cite{Kramer_EffectiveSecrecy2014,endo2014reliability}. Specifically, the latter requires that the distribution of the eavesdropper's observation sequence $\mathbf{Z}$ is asymptotically indistinguishable from some target product measure $q_Z^n$. Fixing a stealth-wiretap code for the WTBC, an analogous GP broadcast channel (GPBC), shown in Fig.~\ref{FIG:WT-GP-BC_intro}(b), is constructed~by:
\begin{enumerate}
\item Converting the eavesdropper's observation $\mathbf{Z}$ to an independently and identically distributed (i.i.d.) state sequence, whose distribution is specified by the WTBC stealth condition;
\item Non-causally revealing the state sequence to the encoder;
\item Setting $p_{Y_1,Y_2|X,Z}$ (the conditional marginal of the WTBC's transition probability, with $Z$ in the role of the state) as the GPBC's transition kernel.
\end{enumerate}
The aforementioned claim relating good sequences of codes for analogous WTBCs and GPBCs remains valid in this extended framework. Through this relation we can capitalize on known GPBC capacity results to derive converse bounds for their analogous WTBCs.


The GPBC has been widely studied in the literature and the capacity region is known for various cases. Notably, the semi-deterministic (SD) GPBC was solved in~\cite{Lapidoth_senideterministic2012}, the physically degraded (PD) GPBC with an informed receiver (i.e., when the stronger receiver also observes the state sequence) was treated in~\cite{Steinberg_Degraded_BC_State2005}, and the cooperative version of this PD-GPBC, where the receivers are connected via a unidirectional noiseless link, was the focus of~\cite{Dikstein_PDBC_Cooperation2016}. The corresponding (SD and PD) WTBCs also received considerable attention in the literature~\cite{Ulukus_WTBC2009,Ulukus_WTBC2013,Yassaee_random_binning2014,Piantanida_External_Eve2015}; however, solutions are known only for some special cases. To the best of our knowledge, the widest framework of DM-WTBCs for which tight secrecy-capacity results are available is due to~\cite{Piantanida_External_Eve2015}, where the regions for the SD-WTBC and the PD-WTBC were derived under the weak-secrecy metric, while assuming that the stochastic receiver is less noisy than the eavesdropper. The coding schemes therein does not rely on this less-noisy property; the converse proofs, however, do.

As compared to WTBCs, GPBCs typically lend themselves to easier converse proofs. This is due to the simpler statistical relations between the random variables in GPBC setups. Specifically, the state $\mathbf{Z}$ being i.i.d. and independent of the messages is crucial for the converse proofs from~\cite{Lapidoth_senideterministic2012,Steinberg_Degraded_BC_State2005,Dikstein_PDBC_Cooperation2016}. In contrast, the WTBC does not share these properties. When $\mathbf{Z}$ is the eavesdropper's observation sequence induced by the code, it is generally not i.i.d., nor is it independent of the message pair. This means that one cannot simply repeat the steps from the converse proof of the analogous GPBC when upper bounding the transmission rates achievable over a WTBC. Nonetheless, our analogy-based proof method enables exploiting the aforementioned GPBC results to characterize the SD- and the PD-WTBC secrecy-capacity regions. As opposed to \cite{Piantanida_External_Eve2015}, while our proof adds a stealth criterion to the WTBC code, it requires no less-noisiness assumptions on the WTBC's outputs. As a natural extension to the base case (WTCs versus GPCs), the obtained WTBC regions are described by the same rate bounds as their GPBC counterparts. The goal of this work is to propose our analogy framework as a new research tool for information-theoretic security and the derived capacity results serve as demonstrations of its usage.

\subsection{Dualities and Analogies in Information Theory}
Connecting seemingly unrelated information-theoretic problems dates back to Shannon's landmark 1959 paper~\cite{Shannon_rate_distortion1959}, where he observed the similarity between channel transmission and lossy source compression. Shannon noticed that two lossy source and channel coding problems that are obtainable from one another by interchanging the roles of their encoder(s) and decoder(s) have dual solutions. Channel and source duality was widely studied ever since, with multiple extensions to multiuser scenarios (a partial list of references is~\cite{cover_duality2002,pradhan_duality2002,yu_duality2002,wang_duality2004,Shirazi_Duality_IEEEI2010,verdu_duality2011,Successive_Refinemen_Permuter2013,Dikstein_MAC_Action_2015,Goldfeld_BC_Cooperation2014}); numerous examples were found that support Shannon's observation. However, a formal definition of this duality remains elusive and, in the absence of a definition, a proof is out of reach. Since there is no mechanism of formal inference in place, channel-source duality is mainly used as a tool to produce educated guesses of the solution of one problem, provided the solution of the other.

An alternative and proven form of duality was established between the Gaussian BC and multiple-access channels (MACs)~\cite{Goldsmith_duality2004}. It was shown that if the channels have corresponding gain and noise parameters, then the capacity-region of one problem can be expressed in terms of the region of the other. This duality was extended to multiple-antenna setups in the followup work~\cite{Goldsmith_duality2003} and was used in~\cite{Goldsmith_duality2003,Tse_duality2003} to find the sum capacity of the multiple-antenna BC. However, the validity of this relation to broader (not necessarily Gaussian) frameworks remains an open question. Analogical relations between different problems were also observed before. In particular, an analogy between Kelly Gambling~\cite{Kelly_gambling1956} and work extraction in statistical physics was the focus of~\cite{Vinkler_duality2016}. 

Perhaps most relevant to our work is~\cite{kibloff2016duality}, which focuses on the relation between GPCs and WTCs. This paper shows that a good random coding ensemble generated by a fixed input distribution for one problem gives rise to a good ensemble for the other (claims in both directions are given). The results assume that the wiretap and GP input distributions, the GP state distribution and the two transition kernels jointly satisfy certain information inequalities. Under these conditions the cross-problem achievability claim holds by generally achieves sub-optimal rates. The paper also identifies a further conditions under which optimality is attained. 
It is noteworthy that the approach from \cite{kibloff2016duality} relies on knowing the single-letter solutions to both problems, which limits the application of this framework to multiuser scenarios. 
Another work that hints on the relation between GPCs and WTCs is ~\cite{liang2009information}, where the secrecy-capacity of a GP-WTC (i.e., a state-dependent WTC with non-causal encoder CSI) was upper bounded by converting it into an enhanced WTC where the transmitter controls the state bound to a type constraint. To the best of our knowledge, the analogy paradigm established here is the first proven relation between the operational setups of the wiretap and GP coding scenarios, which requires no prior knowledge of either solution. Like the aforementioned \emph{proven} duality and analogy relations, it constitutes a research tool that enables inference of results from one problem to the other.

\subsection{Organization}

\par The remainder of this paper is organized as follows. Section~\ref{SEC:preliminaries} provides notation, basic definitions and properties. In Section~\ref{SEC:setup}, we define the WTBC and the GPBC. Section~\ref{SEC:Analogy_explained} explains the analogy between WTCs and GPCs, and illustrates how to use it to prove the converse of the WTC secrecy-capacity (which equals the regular term from \eqref{EQ:WTC_capacity_intro}). The same section also extends the analogy to multiuser broadcasting setups. In Section~\ref{SEC:Results} we state the WTBC secrecy-capacity results derived via the analogy, and cite the past results for the corresponding GPBCs. Proofs are provided in Section~\ref{SEC:proofs}, while Section~\ref{SEC:summary} summarizes the main achievements and insights of this work.


\section{Notations and Preliminaries}\label{SEC:preliminaries}

\par We use the following notations. As is customary, $\mathbb{N}$ is the set of natural numbers (which does not include 0), while $\mathbb{R}$ denotes the reals. We further define $\mathbb{R}_+=\{x\in\mathbb{R}|x\geq 0\}$ and $\mathbb{R}_{++}=\{x\in\mathbb{R}|x> 0\}$. Given two real numbers $a,b$, we denote by $[a\mspace{-3mu}:\mspace{-3mu}b]$ the set of integers $\big\{n\in\mathbb{N}\big| \lceil a\rceil\leq n \leq\lfloor b \rfloor\big\}$. Calligraphic letters denote sets, e.g., $\mathcal{X}$, the complement of $\mathcal{X}$ is denoted by $\mathcal{X}^c$, while $|\mathcal{X}|$ stands for its cardinality. $\mathcal{X}^n$ denotes the $n$-fold Cartesian product of $\mathcal{X}$. An element of $\mathcal{X}^n$ is denoted by $x^n=(x_1,x_2,\ldots,x_n)$; whenever the dimension $n$ is clear from the context, vectors (or sequences) are denoted by boldface letters, e.g., $\mathbf{x}$. For any $\mathcal{A}\subseteq[1:n]$, we use $\mathbf{x}_\mathcal{A}=(x_i)_{i\in\mathcal{A}}$ to denote the substring of $x^n$ defined by $\mathcal{A}$, with respect to the natural ordering of $\mathcal{A}$. For instance, for $1\leq i< j\leq n$, we have $\mathbf{x}_{[i:j]}=(x_i,x_{i+1},\ldots,x_j)$.


Let $\big(\mathcal{X},\mathcal{F},\mathbb{P}\big)$ be a probability space, where $\mathcal{X}$ is the sample space, $\mathcal{F}$ is the $\sigma$-algebra and $\mathbb{P}$ is the probability measure. Random variables over $\big(\mathcal{X},\mathcal{F},\mathbb{P}\big)$ are denoted by uppercase letters, e.g., $X$, with conventions for random vectors similar to those for deterministic sequences. The probability of an event $\mathcal{A}\in\mathcal{F}$ is denoted by $\mathbb{P}(\mathcal{A})$, while $\mathbb{P}(\mathcal{A}\big|\mathcal{B}\mspace{2mu})$ denotes the conditional probability of $\mathcal{A}$ given $\mathcal{B}$. We use $\mathds{1}_\mathcal{A}$ to denote the indicator function of $\mathcal{A}$. The set of all probability mass functions (PMFs) on a finite set $\mathcal{X}$ is denoted by $\mathcal{P}(\mathcal{X})$, i.e., 
\begin{equation}
    \mathcal{P}(\mathcal{X})=\left\{p:\mathcal{X}\to[0,1]\Bigg| \sum_{x\in\mathcal{X}}p(x)=1]\right\}.
\end{equation}
PMFs are denoted by the lowercase letters, such as $p$ or $q$, with a subscript that identifies the random variable and its possible conditioning. For example, for a discrete probability space $\big(\mathcal{X},\mathcal{F},\mathbb{P}\big)$ and two random variables $X$ and $Y$ over that space, we use $p_X$, $p_{X,Y}$ and $p_{X|Y}$ to denote, respectively, the marginal PMF of $X$, the joint PMF of $(X,Y)$ and the conditional PMF of $X$ given $Y$. In particular, $p_{X|Y}$ represents a row stochastic matrix whose elements are given by $p_{X|Y}(x|y)=\mathbb{P}\big(X^{-1}(x)|Y^{-1}(y)\big)$. Expressions such as $p_{X,Y}=p_Xp_{Y|X}$ are to be understood as $p_{X,Y}(x,y)=p_X(x)p_{Y|X}(y|x)$, for all $(x,y)\in\mathcal{X}\times\mathcal{Y}$. Accordingly, when three random variables $X$, $Y$ and $Z$ satisfy $p_{X|Y,Z}=p_{X|Y}$, they form a Markov chain, which we denote by $X- Y - Z$. We omit subscripts if the arguments of a PMF are lowercase versions of the random variables.

For a discrete measurable space $(\mathcal{X},\mathcal{F})$, a PMF $q\in\mathcal{P}(\mathcal{X})$ gives rise to a probability measure on $(\mathcal{X},\mathcal{F})$, which we denote by $\mathbb{P}_q$; accordingly, $\mathbb{P}_q\big(\mathcal{A})=\sum_{x\in\mathcal{A}}q(x)$ for every $\mathcal{A}\in\mathcal{F}$. We use $\mathbb{E}_q$ to denote an expectation taken with respect to $\mathbb{P}_q$. Similarly, we use $H_q$ and $I_q$ to indicate that an entropy or a mutual information is calculated with respect to the PMF $q$. For a sequence of random variable $X^n$, if the entries of $X^n$ are drawn in an i.i.d. manner according to $p_X$, then for every $\mathbf{x}\in\mathcal{X}^n$ we have $p_{X^n}(\mathbf{x})=\prod_{i=1}^np_X(x_i)$ and we write $p_{X^n}(\mathbf{x})=p_X^n(\mathbf{x})$. Similarly, if for every $(\mathbf{x},\mathbf{y})\in\mathcal{X}^n\times\mathcal{Y}^n$ we have $p_{Y^n|X^n}(\mathbf{y}|\mathbf{x})=\prod_{i=1}^np_{Y|X}(y_i|x_i)$, then we write $p_{Y^n|X^n}(\mathbf{y}|\mathbf{x})=p_{Y|X}^n(\mathbf{y}|\mathbf{x})$. The conditional product PMF $p_{Y|X}^n$ given a specific sequence $\mathbf{x}\in\mathcal{X}^n$ is denoted by $p_{Y|X=\mathbf{x}}^n$.

The empirical PMF $\nu_{\mathbf{x}}$ of a sequence $\mathbf{x}\in\mathcal{X}^n$ is
\begin{equation}
	\nu_{\mathbf{x}}(x)\triangleq\frac{N(x|\mathbf{x})}{n},
\end{equation}
where $N(x|\mathbf{x})=\sum_{i=1}^n\mathds{1}_{\{x_i=x\}}$. We use $\mathcal{T}_\epsilon^n(p)$ to denote the set of letter-typical sequences of length $n$ with respect to the PMF $p\in\mathcal{P}(\mathcal{X})$ and the non-negative number $\epsilon$~\cite[Chapter 3]{Massey_Applied}, i.e., we have
\begin{equation}
	\mathcal{T}_\epsilon^n(p)=\Big\{\mathbf{x}\in\mathcal{X}^n\Big|\mspace{5mu}\big|\nu_{\mathbf{x}}(x)-p(x)\big|\leq\epsilon p(x),\ \forall x\in\mathcal{X}\Big\}.
\end{equation}

For a countable sample space $\Omega$ and $p,q\in\mathcal{P}(\Omega)$, the KL divergence between $p$ and $q$ is
\begin{equation}
	\mathsf{D}(p\|q)=\sum_{x\in\supp(p)}p(x)\log\left(\frac{p(x)}{q(x)}\right)\label{EQ:relative_entropy_def_discrete}
\end{equation}
and the \emph{total variation} between them is
\begin{equation}
	\|p-q\|_{\mathsf{TV}}=\frac{1}{2}\sum_{x\in\Omega}\big|p(x)-q(x)\big|=\sum_{\substack{x\in\Omega:\\p(x)>q(x)}}\mspace{-5mu}\big[p(x)-q(x)\big].\label{EQ:total_variation_def_discrete}
\end{equation}
	
Total variation is a distance between probability measures that satisfies the following properties (see, e.g.,~\cite[Property~1]{Cuff_Song_Likelihood2016}). 

\begin{lemma}[Properties of Total Variation]\label{LEMMA:TV_properties}
Let $\Omega$ be a countable sample space and $p,q,r\in\mathcal{P}(\Omega)$. We have:
\begin{enumerate}
    
    \item \underline{Triangle inequality:} $\|p-r\|_{\mathsf{TV}}\leq \|p-q\|_{\mathsf{TV}}+\|q-r\|_{\mathsf{TV}}$.
    
    \item \underline{Difference of Expectations:} If $f:\Omega\to\mathbb{R}$ is a bounded function with $\sup_x|f(x)|\leq b$, then $\big|\mathbb{E}_p f-\mathbb{E}_q f\big|\leq 2b\|p-q\|_\mathsf{TV}$.
    
    \item \underline{Joint / Marginal / Conditional Distributions:} If $\Omega=\mathcal{X}\times\mathcal{Y}$, $p=p_Xp_{Y|X}$ and $q=q_Xq_{Y|X}$, where $p_X,q_X\in\mathcal{P}(\mathcal{X})$ and $p_{Y|X},q_{Y|X}:\mathcal{X}\to\mathcal{P}(\mathcal{Y})$, then
    \begin{enumerate}
        \item $\|p_X-q_X\|_{\mathsf{TV}}\leq \|p-q\|_{\mathsf{TV}}$;
        \item $\|p_Xp_{Y|X}-q_Xp_{Y|X}\|_{\mathsf{TV}}= \|p_X-q_X\|_{\mathsf{TV}}$.
    \end{enumerate}
\end{enumerate}
\end{lemma}

KL divergence dominates total variation through Pinsker's inequality~\cite[Theorem 4.1]{Csiszar_Pinkser_Ineq1967}, implying that for any $p,q\in\mathcal{P}(\Omega)$
\begin{equation}
\|p-q\|_{\mathsf{TV}}\leq \sqrt{\frac{1}{2}\mathsf{D}(p\|q)}.\label{EQ:Pinsker_Inequality}
\end{equation}




The last preliminary result we need states that over finite probability spaces an exponentially decaying total variation dominates the difference between two corresponding mutual information terms. This assertion is formulated in the following lemma. The lemma is stated in a manner that best suits our needs; however, alternative statements~\cite[Lemma 1]{Csiszar_Strong_Secrecy1996} and even better estimates are available in the literature, e.g., in~\cite[Theorem 2]{Zhang_MI_estimation2007}. The proof is omitted.

\begin{lemma}[Total Variation and Mutual Information]\label{LEMMA:MI_continuity}
Let $\mathcal{X}$ and $\mathcal{Y}$ be finite sets (with the obvious $\sigma$-algebras) and let $\{\mu_n\}_{n\in\mathbb{N}}$ and $\{\nu_n\}_{n\in\mathbb{N}}$ be two sequences of probability distributions over $\mathcal{X}^n\times\mathcal{Y}^n$, and denote their marginals on $\mathcal{X}^n$ and $\mathcal{Y}^n$ by $\mu_{X_n}$, $\nu_{X_n}$ and $\mu_{Y_n}$, $\nu_{Y_n}$, respectively. Let $\{\epsilon_n\}_{n\in\mathbf{N}}$ be a sequence of real numbers with $\lim_{n\to\infty}\epsilon_n=0$, such that
\begin{subequations}
\begin{equation}
    \big\|\mu_n-\nu_n\big\|\leq\epsilon_n.\label{EQ:MI_continuity_hypothesis}
\end{equation}
Then, there exists an $n_0\in\mathbb{N}$, such that for any $n>n_0$
\begin{equation}
    \big|I_{\mu_n}(\mathbf{X};\mathbf{Y})-I_{\nu_n}(\mathbf{X};\mathbf{Y})\big|\leq 2n\epsilon_n\log|\mathcal{X}\|\mathcal{Y}|-3\epsilon_n\log\epsilon_n,\label{EQ:MI_continuity_implication}
\end{equation}
\end{subequations}
where $I_{\mu_n}(\mathbf{X};\mathbf{Y})=\mathsf{D}\big(\mu_n\big\|\mu_{X_n}\mu_{Y_n}\big)$ and $I_{\nu_n}(\mathbf{X};\mathbf{Y})=\mathsf{D}\big(\nu_n\big\|\nu_{X_n}\nu_{Y_n}\big)$.

Consequently, if there exists $\beta_1>0$ and an $n_1\in\mathbb{N}$ such that for any $n>n_1$ we have
$\big\|\mu_n-\nu_n\big\|\leq e^{-n\beta_1}$,
then there exists $\beta_2>0$ and an $n_2\in\mathbb{N}$, such that for any $n>n_2$, $\big|I_{\mu_n}(\mathbf{X};\mathbf{Y})-I_{\nu_n}(\mathbf{X};\mathbf{Y})\big|\leq e^{-n\beta_2}$.
\end{lemma}

\section{Problem Setup}\label{SEC:setup}

This section sets up the two problems considered in this work, namely, that of the WTBC and the GPBC.

\subsection{Wiretap Broadcast Channels}\label{SUBSEC:WTBC_def}

Let $\mathcal{X},\ \mathcal{Y}_1,\ \mathcal{Y}_2$ and $\mathcal{Z}$ be finite sets and let $p_{Y_1,Y_2,Z|X}:\mathcal{X}\to\mathcal{P}(\mathcal{Y}_1\times\mathcal{Y}_2\times\mathcal{Z})$ be a transition probability distribution from $\mathcal{X}$ to $\mathcal{Y}_1\times\mathcal{Y}_2\times\mathcal{Z}$. The $\big(\mathcal{X},\mathcal{Y}_1,\mathcal{Y}_2,\mathcal{Z},p_{Y_1,Y_2,Z|X}\big)$ DM-WTBC is illustrated in Fig.~\ref{FIG:WT-GP-BC_intro}(a). The sender chooses a pair of messages $(m_1,m_2)$ uniformly at random from product set $\big[1:2^{nR_1}\big]\times\big[1:2^{nR_2}\big]$ and maps it onto a sequence $\mathbf{x}\in\mathcal{X}^n$ (the mapping may be random). The sequence $\mathbf{x}$ is transmitted over the DM-WTBC with transition probability $p_{Y_1,Y_2,Z|X}$. The output sequences $\mathbf{y}_1\in\mathcal{Y}_1^n$, $\mathbf{y}_2\in\mathcal{Y}_2^n$ and $\mathbf{z}\in\mathcal{Z}^n$ are observed by Receiver 1, Receiver 2 and the eavesdropper, respectively. Based on $\mathbf{y}_j$, $j=1,2$, Receiver $j$ produces an estimate $\hat{m}_j$ of $m_j$. The eavesdropper tries to glean whatever it can about the transmitted messages $(m_1,m_2)$ from $\mathbf{z}$.

\begin{definition}[Classes of WTBCs]
Consider a WTBC $\big(\mathcal{X},\mathcal{Y}_1,\mathcal{Y}_2,\mathcal{Z},p_{Y_1,Y_2,Z|X}\big)$:
\begin{itemize}
\item The channel is called SD if its channel transition distribution factors as $p_{Y_1,Y_2,Z|X}=\mathds{1}_{\{Y_1=y_1(X)\}}p_{Y_2,Z|X}$, where $y_1:\mathcal{X}\to\mathcal{Y}_1$ and $p_{Y_2,Z|X}:\mathcal{X}\to\mathcal{P}(\mathcal{Y}_2\times\mathcal{Z})$.
\item The channel is called PD if its channel transition distribution factors as $p_{Y_1,Y_2,Z|X}=p_{Y_1,Z|X}p_{Y_2|Y_1}$, where $p_{Y_1,Z|X}:\mathcal{X}\to\mathcal{P}(\mathcal{Y}_1\times\mathcal{Z})$ and $p_{Y_2|Y_1}:\mathcal{Y}_1\to\mathcal{P}(\mathcal{Y}_2)$.
\item The channel is said to have an informed receiver if the eavesdropper's output sequence is available to Receiver 1. Formally, replacing the output random variable $Y_1$ with $\tilde{Y}_1=(Y_1,Z)$ reduces the WTBC to have an informed receiver. With a slight abuse of notation, we refer to this setup as the $\big(\mathcal{X},\mathcal{Y}_1,\mathcal{Y}_2,\mathcal{Z},p_{Y_1,Y_2,Z|X}\big)$ PD-WTBC with an informed receiver. 
\end{itemize}
\end{definition}

We proceed with some definitions for general WTBCs and will specialize to the particular instances described above when needed.

\begin{definition}[WTBC Code]\label{DEF:WTBC_code}
	An $(n,R_1,R_2)$-code $c_n$ for the WTBC with a product message set $\mathcal{M}^{(n)}_1\times\mathcal{M}^{(n)}_2$, where for $j=1,2$ we set $\mathcal{M}^{(n)}_1\triangleq[1:2^{nR_j}]$, is a triple of functions $\left(f_n,\phi^{(n)}_1,\phi^{(n)}_2\right)$ such that
	\begin{enumerate}
	    \item $f_n:\mathcal{M}^{(n)}_1\times\mathcal{M}^{(n)}_2\to\mathcal{P}(\mathcal{X}^n)$ is a stochastic encoder;
	    \item $\phi^{(n)}_j:\mathcal{Y}_j^n\to\mathcal{M}^{(n)}_j$ is the decoding function for Receiver $j$, for $j=1,2$.	    
	\end{enumerate}
\end{definition}
	
For any $(n,R_1,R_2)$-code $c_n=\left(f_n,\phi^{(n)}_1,\phi^{(n)}_2\right)$, the induced joint distribution on $\mathcal{M}^{(n)}_1\times\mathcal{M}^{(n)}_2\times\mathcal{X}^n\times\mathcal{Y}_1^n\times\mathcal{Y}_2^n\times\mathcal{Z}^n\times\mathcal{M}^{(n)}_1\times\mathcal{M}^{(n)}_2$ is:
\begin{align*}
	&P^{(c_n)}(m_1,m_2,\mathbf{x},\mathbf{y}_1,\mathbf{y}_2,\mathbf{z},\hat{m}_1,\hat{m}_2)\\
	&=\frac{1}{\big|\mathcal{M}^{(n)}_1\big\|\mathcal{M}^{(n)}_2\big|}f_n(\mathbf{x}|m_1,m_2)p^n_{Y_1,Y_2,Z|X}(\mathbf{y}_1,\mathbf{y}_2,\mathbf{z}|\mathbf{x})\mspace{2mu}\\
	&\quad\quad\quad\quad\quad\quad\quad\quad\quad\quad\quad\quad\quad\times\mathds{1}_{\mspace{-7mu}\bigcap\limits_{j=1,2}\mspace{-5mu}\big\{\hat{m}_j=\phi^{(n)}_j(\mathbf{y}_j)\big\}}.\numberthis\label{EQ:WTBC_induced_PMF}
\end{align*}

Our analogy relies on developing a unified perspective on wiretap and GPCs. We arrive at the desired unification by defining wiretap achievability in a manner slightly different from typical definitions and incorporating a stealth condition \cite{Kramer_EffectiveSecrecy2014,endo2014reliability}. Common definitions require the existence of a sequence of codes that achieves reliability (i.e., a vanishing error probability) and security (e.g., vanishing information leakage and stealth measures). These are two separate requirements that do not project on one another. Namely, a code for the WTBC can be reliable but not secure, secure but not reliable, neither or both.

We take a different approach and merge the reliability and security criteria into a single requirement on the induced distribution in \eqref{EQ:WTBC_induced_PMF} to approximate a certain target measure (e.g., in total variation or rgence). Such definitions have been more frequently used in recent years (see, e.g.,~\cite{Yassaee_fidelity_ISIT2015,Yassaee_random_binning2014}) and seem to originate from~\cite{Watanabe_Converses2014}. Section~\ref{SUBSEC:new_achievability_explained} expounds upon the definition of achievability used herein and shows it is closely related to more standard definitions. 

\begin{definition}[WTBC Achievability]\label{DEF:WTBC_achievability}
A rate pair $(R_1,R_2)\in\mathbb{R}^2_+$ is called achievable if there exists a $\gamma>0$, a probability distribution $q_Z\in\mathcal{P}(\mathcal{Z})$ and a sequence of $(n,R_1,R_2)$-codes $\{c_n\}_{n\in\mathbb{N}}$ such that for any sufficiently large $n$
\begin{align*}
    &\left\|P^{(c_n)}_{M_1,M_2,\hat{M}_1,\hat{M}_2,\mathbf{Z}}-p^{(\mathsf{U})}_{\mathcal{M}^{(n)}_1\times\mathcal{M}^{(n)}_2}\mathds{1}_{\big\{(\hat{M}_1,\hat{M}_2)=(M_1,M_2)\big\}}q_Z^n\right\|_\mathsf{TV}\\&\mspace{380mu}\leq e^{-n\gamma},\numberthis\label{EQ:WTBC_achievability_def}
\end{align*}
where $P^{(c_n)}_{M_1,M_2,\hat{M}_1,\hat{M}_2,\mathbf{Z}}$ is a marginal of the induced joint distribution in \eqref{EQ:WTBC_induced_PMF} and $p^{(\mathsf{U})}_\mathcal{A}$ denotes the uniform distribution over $\mathcal{A}$.
\end{definition}

\begin{remark}[Rate of Convergence]
The exponential rate of convergence in \eqref{EQ:WTBC_achievability_def} is not necessary. Any super-linear convergence rate is sufficient for the claims in this work to go through. The core property we rely on is that sufficiently fast convergence of total variation dominates the difference between associated mutual information terms (Lemma \ref{LEMMA:MI_continuity}). Indeed, Eqn. \eqref{EQ:MI_continuity_implication} shows that so long as $\epsilon_n=o(n)$, the right-hand side (RHS) decays to 0 with $n$ (note that the cardinalities of the alphabets $\mathcal{X}$ and $\mathcal{Y}$ grow at most exponentially fast in this work). 
Thus, to adjust our arguments to the super-linear rate scenario one may replace exponentially decaying terms, such as $e^{-\delta n}$ from \eqref{EQ:WTC_converse_proof_UB2}, with corresponding super-linearly decaying terms. Upon normalizing by $n$ and taking the limit as $n\to\infty$, this has no effect on the final result.
\end{remark}

\begin{definition}[WTBC Secrecy-Capacity Region]
The secrecy-capacity region $\mathcal{C}_\mathsf{WT}(p_{Y_1,Y_2,Z|X})\subseteq\mathbb{R}_+^2$ of a DM-WTBC with transition probability $p_{Y_1,Y_2,Z|X}$ is the convex closure of the set of all achievable rate pairs.
\end{definition}

\begin{remark}[Secrecy and Stealth Conditions]
As seen from our definition of achievability (Definition \ref{DEF:WTBC_achievability}), the considered notion of security requires not only the asymptotic independence of $(M_1,M_2)$ and $\mathbf{Z}$, but also that the distribution of $\mathbf{Z}$ is asymptotically indistinguishable from some product measure $q_Z^n$. The latter is known in the information-theoretic literature as a \emph{stealth} condition \cite{Kramer_EffectiveSecrecy2014,endo2014reliability}, on which we elaborate in the next subsection. We therefore stress the secrecy-capacity region $\mathcal{C}_\mathsf{WT}(p_{Y_1,Y_2,Z|X})$ is with respect to both \emph{secrecy} and \emph{stealth} and as such, it is generally not larger than the corresponding region when the stealth requirement is removed. All of our subsequently presented results account only for when a stealth condition (on top of standard secrecy) is present.
\end{remark}


\subsection{Discussing the Definition of Achievability}\label{SUBSEC:new_achievability_explained}

We discuss the definition of achievability used in this work (Definition~\ref{DEF:WTBC_achievability}), interpret Equation \eqref{EQ:WTBC_achievability_def} and compare our definition to more frequently used notions of achievability. 

Equation \eqref{EQ:WTBC_achievability_def} means that a good sequence of codes induces a joint distribution (see \eqref{EQ:WTBC_induced_PMF}) whose $\big(M_1,M_2,\hat{M}_1,\hat{M}_2,\mathbf{Z}\big)$-marginal is asymptotically indistinguishable from the target measure $p^{(\mathsf{U})}_{\mathcal{M}^{(n)}_1\times\mathcal{M}^{(n)}_2}\mathds{1}_{\big\{(\hat{M}_1,\hat{M}_2)=(M_1,M_2)\big\}}q_Z^n$. To interpret this requirement, observe that:
\begin{enumerate}
    
    \item Under the target measure, $(\hat{M}_1,\hat{M}_2)=(M_1,M_2)$ almost surely. This corresponds to reliable decoding by both receivers. 
    
    \item Under the target measure, the conditional distribution of $\mathbf{Z}$ given $(M_1,M_2,\hat{M}_1,\hat{M}_2)$ is the product measure $q_Z^n$. Thus, as $n$ grows, the induced distribution of $(M_1,M_2,\mathbf{Z})$ approaches  $p^{(\mathsf{U})}_{\mathcal{M}^{(n)}_1\times\mathcal{M}^{(n)}_2}\times q_Z^n$, which accounts for both \emph{security} and \emph{stealth}~\cite{Kramer_EffectiveSecrecy2014,endo2014reliability}.\footnote{Formally, stealth refers to a vanishing divergence between the induced $P^{(c_n)}_{\mathbf{Z}}$ and some target product $q_Z^n$. Different divergences may be considered, such as the KL divergence $\mathsf{D}\big(P^{(c_n)}_\mathbf{Z}\big\|q_Z^n\big)\xrightarrow[n\to\infty]{}0$ or total variation $\big\|P^{(c_n)}_\mathbf{Z}-q_Z^n\big\|_{\mathsf{TV}}\xrightarrow[n\to\infty]{}0$.} Intuitively, this means that a good sequence of codes not only secures the transmission, but also makes it impossible for the eavesdropper to determine whether communication at all occurred (or was the channel fed with random noise i.i.d. according to some $q_X\in\mathcal{P}(\mathcal{X})$ with $\sum_{x,y_1,y_2}q_X(x)p_{Y_1,Y_2,Z|X}(y_1,y_2,z|x)=q_Z(z)$, for all $z\in\mathcal{Z}$).

\end{enumerate}

While similar definitions of achievability were used in~\cite{Watanabe_Converses2014,Yassaee_random_binning2014,Yassaee_fidelity_ISIT2015}, a more standard definition was employed in~\cite{Kramer_EffectiveSecrecy2014,endo2014reliability}. In that work, reliability was defined through a vanishing probability of error requirement, while security was measured by the \emph{effective secrecy} metric. Adapting the definitions from~\cite{Kramer_EffectiveSecrecy2014,endo2014reliability} to the WTBC considered here gives rise to the following notion of achievability (to differentiate from Definition~\ref{DEF:WTBC_achievability}, we refer to this as \emph{EF-achievability}, where `EF' stands for `effective secrecy').

\begin{definition}[EF-Achievability]\label{DEF:WTBC_achievability_classic}
A rate pair $(R_1,R_2)\in\mathbb{R}^2_+$ is called EF-achievable if there exists a probability distribution $q_Z\in\mathcal{P}(\mathcal{Z})$ and a sequence of $(n,R_1,R_2)$-codes $\{c_n\}_{n\in\mathbb{N}}$ such that
\begin{subequations}
\begin{align}
\mathbb{P}_{P^{(c_n)}}\Big(\big(\hat{M}_1,\hat{M}_2\big)\neq(M_1,M_2)\Big)&\xrightarrow[n\to\infty]{}0\label{EQ:WTBC_cachievability_reliable}\\
\mathsf{D}\Big(P^{(c_n)}_{M_1,M_2,\mathbf{Z}}\Big\|p^{(\mathsf{U})}_{\mathcal{M}^{(n)}_1\times\mathcal{M}^{(n)}_2}q_Z^n \Big)&\xrightarrow[n\to\infty]{}0.\label{EQ:WTBC_cachievability_secure}
\end{align}\label{EQ:WTBC_cachievability}%
\end{subequations}
\end{definition}

Convergence to zero of the effective secrecy metric in \eqref{EQ:WTBC_cachievability_secure} guarantees both \emph{strong secrecy} and \emph{stealth}. Strong secrecy refers to a vanishing mutual information between the confidential message and the eavesdropper's observation, i.e., $I_{P^{(c_n)}}(M_1,M_2;\mathbf{Z})\to 0$ as $n\to\infty$. Stealth is quantified by means of KL divergence $\mathsf{D}\Big(P^{(c_n)}_{\mathbf{Z}}\Big\|q_Z^n\Big)$ between the induced distribution $P^{(c_n)}_{\mathbf{Z}}$ and the product measure $q_Z^n$. 
Noting that the KL divergence in \eqref{EQ:WTBC_cachievability_secure} factors~as
\begin{align*}
\mathsf{D}\Big(&P^{(c_n)}_{M_1,M_2,\mathbf{Z}}\Big\|p^{(\mathsf{U})}_{\mathcal{M}^{(n)}_1\times\mathcal{M}^{(n)}_2}q_Z^n \Big)\\&=\underbrace{I_{P^{(c_n)}}(M_1,M_2;\mathbf{Z})}_{\mbox{Strong secrecy measure}}+\underbrace{\mathsf{D}\Big(P^{(c_n)}_{\mathbf{Z}}\big\|q_Z^n\Big)}_{\mbox{Stealth measure}}\numberthis\label{EQ:effective_secrecy_def}
\end{align*}
we see that \eqref{EQ:WTBC_cachievability_secure} indeed implies strong secrecy and stealth.

There is a close relation between the two notions of achievability from Definitions~\ref{DEF:WTBC_achievability} and~\ref{DEF:WTBC_achievability_classic}. This is formalized in the following proposition. 

\begin{proposition}[Relation between Definitions~\ref{DEF:WTBC_achievability} and~\ref{DEF:WTBC_achievability_classic}]\label{PROP:Achievability_Relation}
Let $\{c_n\}_{n\in\mathbb{N}}$ be a sequence of $(n,R_1,R_2)$-codes for the DM-WTBC $\big(\mathcal{X},\mathcal{Y}_1,\mathcal{Y}_2,\mathcal{Z},p_{Y_1,Y_2,Z|X}\big)$. The following implications hold:
\begin{enumerate}
    \item If there exist $\gamma_1>0$ and $q_Z\in\mathcal{P}(\mathcal{Z})$ such that for any $n$ sufficiently large 
    \begin{align*}
    &\mspace{-15mu}\left\|\mspace{-1mu}P^{(c_n)}_{\mspace{-2mu}M_1\mspace{-1mu},\mspace{-1mu}M_2\mspace{-1mu},\mspace{-1mu}\hat{M}_1\mspace{-1mu},\mspace{-1mu}\hat{M}_2\mspace{-1mu},\mspace{-1mu}\mathbf{Z}}\mspace{-6mu}-\mspace{-3mu}p^{(\mathsf{U})}_{\mspace{-3mu}\mathcal{M}^{\mspace{-1.5mu}(\mspace{-1mu}n\mspace{-1mu})}_1\mspace{-2mu}\times\mspace{-2mu}\mathcal{M}^{\mspace{-1.5mu}(\mspace{-1mu}n\mspace{-1mu})}_2}\mspace{-2mu}\mathds{1}_{\mspace{-2mu}\big\{\mspace{-3mu}(\mspace{-1mu}\hat{M}_1\mspace{-1mu},\mspace{-1mu}\hat{M}_2\mspace{-1mu})=(\mspace{-1mu}M_1\mspace{-1mu},\mspace{-1mu}M_2\mspace{-1mu})\mspace{-3mu}\big\}}\mspace{-2mu}q_Z^n\mspace{-1mu}\right\|_\mathsf{TV}\\&\quad\quad\quad\quad\quad\quad\quad\quad\quad\quad\quad\quad\quad\quad\quad\quad\quad\quad\leq e^{-n\gamma_1},\numberthis\label{EQ:WTBC_achievability_exponential}
    \end{align*}
    then  $\{c_n\}_{n\in\mathbb{N}}$ satisfies \eqref{EQ:WTBC_cachievability}.
    \vspace{2mm}
    \item If there exist $\gamma_2>0$ and $q_Z\in\mathcal{P}(\mathcal{Z})$ such that for any $n$ sufficiently large 
    \begin{subequations}
    \begin{align}
    \mathbb{P}_{P^{(c_n)}}\Big(\big(\hat{M}_1,\hat{M}_2\big)\neq(M_1,M_2)\Big)&\leq e^{-n\gamma_2}\label{EQ:WTBC_cachievability_reliable_exp}\\
    \mathsf{D}\Big(P^{(c_n)}_{M_1,M_2,\mathbf{Z}}\Big\|p^{(\mathsf{U})}_{\mathcal{M}^{(n)}_1\times\mathcal{M}^{(n)}_2}q_Z^n \Big)&\leq e^{-n\gamma_2}\label{EQ:WTBC_cachievability_efsecrecy},
    \end{align}\label{EQ:WTBC_cachievability_exponential}%
    \end{subequations}
    then $\{c_n\}_{n\in\mathbb{N}}$ satisfies \eqref{EQ:WTBC_achievability_exponential} for some $\gamma_1>0$ and the same $q_Z$.
\end{enumerate}
\end{proposition}

The proof of Proposition~\ref{PROP:Achievability_Relation} is standard and therefore omitted. It relies on Pinsker's inequality and~\cite[Lemma 2.7]{Csiszar_Korner_Book2011}. The proposition says that an achievable rate pair $(R_1,R_2)$ (with respect to Definition~\ref{DEF:WTBC_achievability}) is always EF-achievable (with respect to Definition~\ref{DEF:WTBC_achievability_classic}). Conversely, given an EF-achievable rate pair $(R_1,R_2)$ with a sequence of codes that produces an exponential decay both in \eqref{EQ:WTBC_cachievability_reliable} and \eqref{EQ:WTBC_cachievability_secure}, we have that $(R_1,R_2)$ is also achievable. The two definitions are therefore equivalent, provided that the aforementioned quantities converge to 0 exponentially fast with~$n$.

\begin{remark}[Natural Sufficient Conditions] We note that the exponential convergence in \eqref{EQ:WTBC_achievability_exponential} and \eqref{EQ:WTBC_cachievability_exponential} is a rather standard feature. The direct proofs of our main secrecy-capacity theorems establish the exponential decay of the total variation in \eqref{EQ:WTBC_achievability_exponential} (see Section \ref{SEC:proofs}). From a broader view, i.i.d. random coding ensembles produce an exponential decay of the expected error probability and the effective secrecy metric (see, e.g., \cite{hou2013informational,Kramer_EffectiveSecrecy2014,Goldfeld_WTCII_semantic2015}). This means that the specific code extracted from the ensemble would also exhibit exponential convergence rates of these two quantities. To see this, let $\big\{\mathsf{C}_n\big\}_{n\in\mathbb{N}}$ be a random coding ensemble, and $\mathsf{P}_\mathsf{e}(c_n)$ and $\mathsf{D}(c_n)$ be the error probability and the effective secrecy metric, respectively (the exact communication scenario is of no consequence here). Assume that there exists $\delta>0$ such that for sufficiently large $n$ we have
$\max\Big\{\mathbb{E}\mathsf{P}_\mathsf{e}(\mathsf{C}_n),\mathbb{E}\mathsf{D}(\mathsf{C}_n)\Big\}\leq e^{-n\delta}$. For every such $n$, the union bound and Markov's inequality give
\begin{align*}
    &\mathbb{P}\bigg(\left\{\mathsf{P}_\mathsf{e}(\mathsf{C}_n)>e^{-n\frac{\delta}{2}}\right\}\cup\left\{\mathsf{D}(\mathsf{C}_n)>e^{-n\frac{\delta}{2}}\right\}\bigg)\\
    &\quad\quad\quad\quad\quad\quad\quad\quad\quad\quad\quad\quad\leq e^{n\frac{\delta}{2}}\mathbb{E}\mathsf{P}_\mathsf{e}(\mathsf{C}_n)+e^{n\frac{\delta}{2}}\mathbb{E}\mathsf{D}(\mathsf{C}_n)\\
    &\quad\quad\quad\quad\quad\quad\quad\quad\quad\quad\quad\quad\leq 2e^{-n\frac{\delta}{2}}.\numberthis
\end{align*}
Thus, for each large enough $n$ there exists a block-code $c'_n$ with an exponentially decaying $\mathsf{P}_\mathsf{e}(c'_n)$ and $\mathsf{D}(c'_n)$. 

\end{remark}


\subsection{Gelfand-Pinsker Broadcast Channels}\label{SUBSEC:GPBC_def}

We derive single-letter characterizations of the secrecy-capacity region of some WTBCs based on the results for analogous GPBCs. The capacity regions of the SD-GPBC and the PD-GPBC with an informed receiver were found in~\cite[Theorem 1]{Lapidoth_senideterministic2012} and~\cite[Theorem 3]{Steinberg_Degraded_BC_State2005}, respectively. These results play a key role in the converse proofs of Theorems~\ref{TM:WTBC_capacity} and~\ref{TM:PD_WTBC_capacity}. We next formally define the GPBC setup. 

Let $\mathcal{Z},\ \mathcal{X},\ \mathcal{Y}_1$ and $\mathcal{Y}_2$ be finite sets,  $q_Z\in\mathcal{P}(\mathcal{Z})$ and $q_{Y_1,Y_2|X,Z}:\mathcal{X}\times\mathcal{Z}\to\mathcal{P}(\mathcal{Y}_1\times\mathcal{Y}_2)$. The $\big(\mathcal{Z},\mathcal{X},\mathcal{Y}_1,\mathcal{Y}_2,q_Z,q_{Y_1,Y_2|X,Z}\big)$ DM-GPBC is shown in Fig.~\ref{FIG:WT-GP-BC_intro}(b). Some particular classes of interest are as follows.

\begin{definition}[Classes of GPBCs]
Consider a GPBC $\big(\mathcal{Z},\mathcal{X},\mathcal{Y}_1,\mathcal{Y}_2,q_Z,q_{Y_1,Y_2|X,Z}\big)$:
\begin{itemize}
\item The channel is called SD if its channel transition distribution factors as $q_{Y_1,Y_2|X,Z}=\mathds{1}_{\{Y_1=y_1(X,Z)\}}q_{Y_2|X,Z}$, where $y_1:\mathcal{X}\times\mathcal{Z}\to\mathcal{Y}_1$ and $q_{Y_2|X,Z}:\mathcal{X}\times\mathcal{Z}\to\mathcal{P}(\mathcal{Y}_2)$.
\item The channel is called PD if its channel transition distribution factors as $q_{Y_1,Y_2|X,Z}=q_{Y_1|X,Z}q_{Y_2|Y_1}$, where $q_{Y_1|X,Z}:\mathcal{X}\times\mathcal{Z}\to\mathcal{P}(\mathcal{Y}_1)$ and $q_{Y_2|Y_1}:\mathcal{Y}_1\to\mathcal{P}(\mathcal{Y}_2)$.
\item The channel is said to have an informed receiver if the state sequence is available to Receiver 1. 
\end{itemize}
\end{definition}

As in Section~\ref{SUBSEC:WTBC_def}, we define codes, achievability and capacity for arbitrary (not necessarily SD) DM-GPBCs, and will specialize to the above instances when necessary.

\begin{definition}[GPBC Code]\label{DEF:GPBC_code}
	An $(n,R_1,R_2)$-code $b_n$ for the GPBC with a product message set $\mathcal{M}^{(n)}_1\times\mathcal{M}^{(n)}_2$, where $\mathcal{M}^{(n)}_j\triangleq[1:2^{nR_j}]$ for $j=1,2$, is a triple of functions $\left(g_n,\psi^{(n)}_1,\psi^{(n)}_2\right)$ such that
	\begin{enumerate}
	    \item $g_n:\mathcal{Z}^n\times\mathcal{M}^{(n)}_1\times\mathcal{M}^{(n)}_2\to\mathcal{P}(\mathcal{X}^n)$ is a stochastic encoder;
	    \item $\psi^{(n)}_j:\mathcal{Y}_j^n\to\mathcal{M}^{(n)}_j$ is the decoding function for Receiver $j$, for $j=1,2$.	    
	\end{enumerate}
\end{definition}
	
An $(n,R_1,R_2)$-code $b_n=\left(g_n,\psi^{(n)}_1,\psi^{(n)}_2\right)$ induces a joint distribution over $\mathcal{Z}^n\times\mathcal{M}^{(n)}_1\times\mathcal{M}^{(n)}_2\times\mathcal{X}^n\times\mathcal{Y}_1^n\times\mathcal{Y}_2^n\times\mathcal{M}^{(n)}_1\times\mathcal{M}^{(n)}_2$ that is given by
\begin{align*}
	Q^{(b_n)}&(\mathbf{z},m_1,m_2,\mathbf{x},\mathbf{y}_1,\mathbf{y}_2,\hat{m}_1,\hat{m}_2)\\\mspace{-17mu}&=q_Z^n(\mathbf{z})\frac{1}{\big|\mathcal{M}^{(n)}_1\big\|\mathcal{M}^{(n)}_2\big|}g_n(\mathbf{x}|\mathbf{z},m_1,m_2)\\&\mspace{-17mu}\quad\quad\times q^n_{Y_1,Y_2|X,Z}(\mathbf{y}_1,\mathbf{y}_2|\mathbf{x},\mathbf{z})\mathds{1}_{\mspace{-6mu}\bigcap\limits_{j=1,2}\mspace{-3mu}\big\{\hat{m}_j=\psi^{(n)}_j(\mathbf{y}_j)\big\}}.\numberthis\label{EQ:GPBC_induced_PMF}
\end{align*}

To enable the use of previous capacity region characterizations, we adhere to the (standard) definition of achievability from~\cite{Steinberg_Degraded_BC_State2005,Lapidoth_senideterministic2012,Dikstein_PDBC_Cooperation2016}. Thus, the error probability associated with an $(n,R_1,R_2)$-code $b_n$ is the probability of the event $\Big\{\big(\hat{M}_1,\hat{M}_2\big)\neq(M_1,M_2)\Big\}$ under the measure induced by the PMF $Q^{(b_n)}$ in \eqref{EQ:GPBC_induced_PMF}. Namely, denoting the probability of error by $\mathsf{P}_\mathsf{e}(b_n)$, we have
$\mathsf{P}_\mathsf{e}(b_n)\triangleq\mathbb{P}_{Q^{(b_n)}}\Big(\big(\hat{M}_1,\hat{M}_2\big)\neq(M_1,M_2)\Big)$. 
    
\begin{definition}[GPBC Achievability]\label{DEF:GPBC_achievability}
A rate pair $(R_1,R_2)\in\mathbb{R}^2_+$ is called achievable if there exists a sequence of $(n,R_1,R_2)$-codes $\{b_n\}_{n\in\mathbb{N}}$ such that $\mathsf{P}_\mathsf{e}(b_n)\xrightarrow[n\to\infty]{}0$.
\end{definition}

\begin{definition}[GPBC Capacity Region]
The capacity region $\mathcal{C}_\mathsf{GP}(q_Z,q_{Y_1,Y_2|X,Z})\subseteq\mathbb{R}_+^2$ of a DM-GPBC with state distribution $q_Z$ and transition probability $q_{Y_1,Y_2|X,Z}$ is the convex closure of the set of all achievable rate pairs. 
\end{definition}



\section{Wiretap and Gelfand-Pinsker Analogy}\label{SEC:Analogy_explained}

The main idea behind the WTC-GPC analogy is rephrasing the two problems to get a unified perspective on them both. Defining achievability in terms of the \emph{induced joint distribution} converging to an appropriate \emph{target measure} under which the performance criteria (e.g., reliability, security and stealth) are satisfied does just that. Consequently, one can show that a sequence of stealth-wiretap codes of a give rate attaining reliability, security and stealth induces a reliable sequence of codes (of the same rate) for a corresponding GPC. Then, one may further deduce that any upper bound on the capacity of this GPC also serves as an upper bound on the achievable wiretap rate we have started from. 

We next describe the analogy principle for the base case of the classic wiretap and GPCs. As a first simple example, the analogy is used to derive a converse proof for the WTC's secrecy-capacity theorem (when a stealth requirement is present; see, e.g., \cite[Theorem 1]{Kramer_EffectiveSecrecy2014}). Then, we outline extensions of this idea to multiuser (namely, broadcasting) scenarios. These extension are used to prove the main secrecy-capacity results of this work that are stated in Section~\ref{SEC:Results}.


\subsection{Base Case - WTC and GPC Analogy}\label{SUBSEC:base_case}


\subsubsection{\underline{A Unified Perspective}}

The WTC and the GPC are related through the fact that their target joint distributions share the same structure. To see this, consider the classic WTC with transition probability $p_{Y,Z|X}:\mathcal{X}\to\mathcal{P}(\mathcal{Y}\times\mathcal{Z})$, for which achievability is defined similarly to Definition~\ref{DEF:WTBC_achievability}, and the point-to-point GPC with state distribution $q_Z\in\mathcal{P}(\mathcal{Z})$ and channel transition probability $q_{Y|X,Z}:\mathcal{X}\times\mathcal{Z}\to\mathcal{P}(\mathcal{Y})$.\footnote{These two problems are special cases of their broadcast versions defined in Sections~\ref{SUBSEC:WTBC_def} and~\ref{SUBSEC:GPBC_def}, respectively. Setting $R_2=0$, $Y_2=0$ and relabeling $\mathcal{M}_1^{(n)}$, $R_1$ and $Y_1$ as $\mathcal{M}_n$, $R$ and $Y$, respectively, reduces the definitions of the broadcast versions to the point-to-point scenario.} The joint distribution induced by an $(n,R)$-code $c_n=(f_n,\phi_n)$ for the wiretap channel is (see \eqref{EQ:WTBC_induced_PMF})
\begin{align*}
	\tilde{P}^{(c_n)}&(m,\mathbf{x},\mathbf{y},\mathbf{z},\hat{m})\\&=\frac{1}{|\mathcal{M}_n|}f_n(\mathbf{x}|m)p^n_{Y,Z|X}(\mathbf{y},\mathbf{z}|\mathbf{x})\mathds{1}_{\big\{\hat{m}=\phi_n(\mathbf{y})\big\}},\numberthis\label{EQ:WTC_induced_PMF}
\end{align*}
while the induced joint distribution for the GPC with respect to an $(n,R)$-code $b_n=(g_n,\psi_n)$ is (see \eqref{EQ:GPBC_induced_PMF})
\begin{align*}
	&\tilde{Q}^{(b_n)}(\mathbf{z},m,\mathbf{x},\mathbf{y},\hat{m})\\&=q_Z^n(\mathbf{z})\frac{1}{|\mathcal{M}_n|}g_n(\mathbf{x}|\mathbf{z},m)q^n_{Y|X,Z}(\mathbf{y}|\mathbf{x},\mathbf{z})\mathds{1}_{\big\{\hat{m}=\psi_n(\mathbf{y})\big\}}.\numberthis\label{EQ:GPC_induced_PMF}
\end{align*}

With respect to Definition~\ref{DEF:WTBC_achievability}, a rate $R$ is achievable for the WTC if there exist a sequence of $(n,R)$-codes $\{c_n\}_{n\in\mathbb{N}}$ and a distribution $q_Z\in\mathcal{P}(\mathcal{Z})$, such that
\begin{equation}
    \left\|\tilde{P}^{(c_n)}_{M,\hat{M},\mathbf{Z}}-p^{(\mathsf{U})}_{\mathcal{M}_n}\mathds{1}_{\{\hat{M}=M\}}q_Z^n\right\|_\mathsf{TV}\xrightarrow[n\to\infty]{}0,\label{EQ:WTC_achievability_def}
\end{equation}
exponentially fast with $n$, where $\tilde{P}^{(c_n)}_{M,\hat{M},\mathbf{Z}}$ is the $(M,\hat{M},\mathbf{Z})$-marginal of $\tilde{P}^{(c_n)}$ in \eqref{EQ:WTC_induced_PMF}. 

For the GPC, achievability is defined by adapting Definition~\ref{DEF:GPBC_achievability} to the point-to-point case. Namely, $R\in\mathbb{R}_+$ is achievable if there exists a sequence of $(n,R)$-codes $\{b_n\}_{n\in\mathbb{N}}$ for the GPC such that
\begin{equation}
    \mathsf{P}_\mathsf{e}(b_n)\triangleq\mathbb{P}_{\tilde{Q}^{(b_n)}}\big(\hat{M}\neq M\big) \xrightarrow[n\to\infty]{}0.\label{EQ:GPC_achievability_error_prob}
\end{equation}
Now, observe that
\begin{align*}
    &\mathbb{P}_{\tilde{Q}^{(b_n)}}\big(\hat{M}\neq M\Big)\\
    &=\sum_{\substack{m,\hat{m}\in\mathcal{M}_n:\\
    m\neq\hat{m}}} \tilde{Q}^{(b_n)}(m,\hat{m})\\
    &=\sum_{\substack{m,\hat{m}\in\mathcal{M}_n:\\
    m\neq\hat{m}}} \left[\tilde{Q}^{(b_n)}(m,\hat{m})-p^{(\mathsf{U})}_{\mathcal{M}_n}(m)\mathds{1}_{\{\hat{m}=m\}}\right]\\
    &\stackrel{(a)}=\sum_{\substack{m,\hat{m}\in\mathcal{M}_n:\\
    \tilde{Q}^{(b_n)}(m,\hat{m})> p^{(\mathsf{U})}_{\mathcal{M}_n}(m)\mathds{1}_{\{\hat{m}=m\}}}} \mspace{-45mu}\left[\tilde{Q}^{(b_n)}(m,\hat{m})-p^{(\mathsf{U})}_{\mathcal{M}_n}(m)\mathds{1}_{\{\hat{m}=m\}}\right]\\
    &\stackrel{(b)}=\left\|\tilde{Q}^{(b_n)}_{M,\hat{M}}-p^{(\mathsf{U})}_{\mathcal{M}_n}\mathds{1}_{\{\hat{M}=M\}}\right\|_\mathsf{TV}\numberthis\label{EQ:GPC_Relation_TV_Pe}
\end{align*}
where (a) is since $m\neq\hat{m}$ if and only if $\tilde{Q}^{(b_n)}(m,\hat{m})\geq p^{(\mathsf{U})}_{\mathcal{M}_n}(m)\mathds{1}_{\{\hat{m}=m\}}$ and because $(m,\hat{m})$ pairs for which we have an equality do not contribute to the summation; (b) is by the second equality in \eqref{EQ:total_variation_def_discrete}. If the error probability decays exponentially fast, then by Pinsker's inequality we have\footnote{The converse claim is also true: convergence to zero of the total variation in \eqref{EQ:GPC_achievability_def} implies $\mathsf{P}_\mathsf{e}(b_n)\xrightarrow[n\to\infty]{}0$. This is an immediate consequence of \eqref{EQ:GPC_Relation_TV_Pe} and Property (3-a) from Lemma~\ref{LEMMA:TV_properties}.}
\begin{equation}
    \left\|\tilde{Q}^{(c_n)}_{M,\hat{M},\mathbf{Z}}-p^{(\mathsf{U})}_{\mathcal{M}_n}\mathds{1}_{\{\hat{M}=M\}}q_Z^n\right\|_\mathsf{TV}\xrightarrow[n\to\infty]{}0,\label{EQ:GPC_achievability_def}
\end{equation}
with an exponential convergence rate.


Having \eqref{EQ:WTC_achievability_def} and \eqref{EQ:GPC_achievability_def}, it is evident that although each problem has its own induced joint distribution, their target measures share the same structure. In both problems, a ``good" sequence of codes induces a sequence of distributions ($\big\{\tilde{P}^{(c_n)}\big\}_{n\in\mathbb{N}}$ or $\big\{\tilde{Q}^{(b_n)}\big\}_{n\in\mathbb{N}}$ for the WTC or the GPC, respectively) that approximates a target distribution where: (i) $M=\hat{M}$ almost surely; (ii) $\mathbf{Z}$ is independent of $M$; and (iii) the entries of $\mathbf{Z}$ are i.i.d. The first item is a consequence of the reliability requirement in both problems. For the second and third items, note that $\mathbf{Z}$ having i.i.d. entries that are independent of $M$ is the security and stealth requirements in the WTC scenario. For the GPC, however, these statistical relations are a part of the problem definition. 

\vspace{2mm}

\subsubsection{\underline{Constructing Analogous Models and Codes}}\label{SUBSUBSEC:analogy_intro}

\begin{figure*}[!t]
	\begin{center}
	    \begin{psfrags}
	        \psfragscanon
	        \psfrag{A}[][][1]{$M$}
	        \psfrag{J}[][][1]{$M$}
	        \psfrag{B}[][][1]{$f_n$}
	        \psfrag{K}[][][1]{$g_n$}
	        \psfrag{C}[][][1]{\ \ \ $\mathbf{X}$}
	        \psfrag{L}[][][1]{\ \ \ $\mathbf{X}$}
	        \psfrag{D}[][][1]{\ \ \ \ \ \ \ $p^n_{Y,Z|X}$}
	        \psfrag{M}[][][1]{\ \ \ \ \ \ \ $p^n_{Y|X,Z}$}
	        \psfrag{E}[][][1]{\ \ \ \ $\mathbf{Y}$}
	        \psfrag{N}[][][1]{\ \ \ \ $\mathbf{Y}$}
	        \psfrag{F}[][][1]{$\phi_n$}
	        \psfrag{O}[][][1]{$\psi_n$}
	        \psfrag{G}[][][1]{\ \ $\hat{M}$}
	        \psfrag{P}[][][1]{\ \ $\hat{M}$}
            \psfrag{H}[][][1]{\ \ \ \ \ \ \ \ \ \ \ \ \ \ \ \ \ \ ${\color{red!65!black}\mathbf{Z}\sim P^{(c_n)}_{\mathbf{Z}|M}\approx q_Z^n}$}
            \psfrag{Z}[][][1]{\ \ \ \ \ \ \ \ \ \ \ \quad\quad  ${\color{red!65!black}\mathbf{Z}\sim P^{(c_n)}_{\mathbf{Z}|M}\approx q_Z^n}$}
	        \psfrag{R}[][][1]{\ \ \ ${\color{red!65!black}q_Z^n}$}
	        \psfrag{Q}[][][1]{${\color{red!65!black}\mathbf{Z}}$}
	        \psfrag{X}[][][1]{$C_\mathsf{WT}=\max\limits_{p_{U,X}}\Big[I(U;Y)-I(U;Z)\Big]$}
	        \psfrag{Y}[][][1]{$C_\mathsf{GP}=\max\limits_{q_{U,X|Z}}\Big[I(U;Y)-I(U;Z)\Big]$}
	        \psfrag{U}[][][1]{$C_\mathsf{WT}^{(\mathsf{IR})}=\max\limits_{p_X}I(X;Y|Z)$}
	        \psfrag{V}[][][1]{$C_\mathsf{GP}^{(\mathsf{IR})}=\max\limits_{q_{X|Z}}I(X;Y|Z)$}
	        \subfloat[]{\includegraphics[scale = .35]{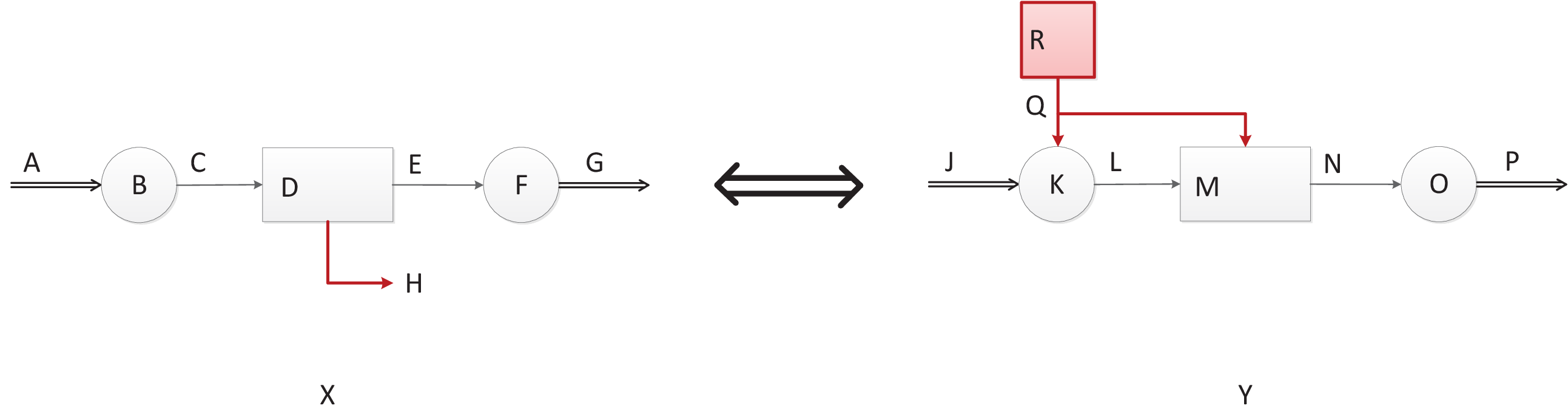}}\\\vspace{3mm}
	        \subfloat[]{\includegraphics[scale = .35]{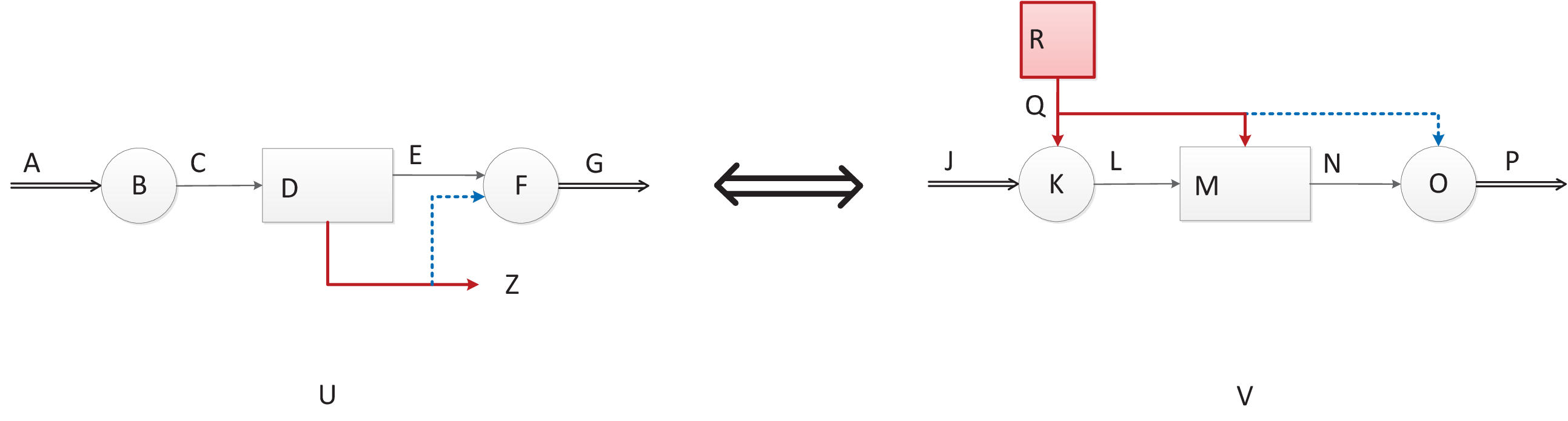}}
	        \caption{WTC and GPC analogy: (a) The GPC is obtained from the WTC by replacing the eavesdropper (whose distribution of the channel observation given the message $P^{(c_n)}_{\mathbf{Z}|M}$ asymptotically approaches the product measure $q_Z^n$) with a state sequence $\mathbf{Z}\sim q_Z^n$, revealing $\mathbf{Z}$ to the encoder and setting the state-dependent channel $p_{Y|X,Z}$ as the conditional marginal distribution of the WTC's transition probability $p_{Y,Z|X}$; (b) If the WTC receiver also observes $\mathbf{Z}$, then the receiver of the analogous GPC is informed of the state.}\label{FIG:WT-GP}
	        \psfragscanoff
	    \end{psfrags}
	\end{center}
\end{figure*}

Equipped with the above perspective, we next show that good stealth-wiretap codes induce good codes for a corresponding GPC. To formalize this statement, we first describe how an \emph{analogous} GPC is constructed from a given WTC and a good sequence of codes for it. Let $\big(\mathcal{X},\mathcal{Y},\mathcal{Z},p_{Y,Z|X}\big)$ be a WTC with a target (product) distribution $q^n_Z$ for the eavesdropper's observation. 
The analogous GPC is obtained by:
\begin{enumerate}
    \item Replacing the eavesdropper of the WTC with a state sequence $\mathbf{Z}$ distributed according to $q^n_Z$;
    \item Revealing $\mathbf{Z}$ in a non-causal manner to the encoder;
    \item Setting the state-dependent channel to $p_{Y|X,Z}$ (the conditional marginal of the WTC's transition probability) with $Z$ in the role of the state.
\end{enumerate}
These steps convert the WTC and the given sequence of codes into a $\big(\mathcal{Z},\mathcal{X},\mathcal{Y},q_Z,p_{Y|X,Z}\big)$ GPC. If the WTC's receiver observes both $\mathbf{Y}$ and $\mathbf{Z}$, then the receiver of the analogous GPC has access to the state sequence $\mathbf{Z}$ (see Fig.~\ref{FIG:WT-GP} for an illustration). Notably, the capacity formulas of two analogous models are given as an optimization (over different domains) of the same information measures.




\begin{proposition}[From Stealth-Wiretap Codes to GP Codes]\label{PROP:WTGP_codes_relations}
Let $\mathcal{X},\ \mathcal{Y}$ and $\mathcal{Z}$ be finite sets and consider a $\big(\mathcal{X},\mathcal{Y},\mathcal{Z},p_{Y,Z|X}\big)$ WTC. Let $R\in\mathbb{R}_+$ be an achievable rate for the WTC, with a corresponding sequence of $(n,R)$-codes $\{c_n\}_{n\in\mathbb{N}}$, where  $c_n=(f_n,\phi_n)$ for each $n\in\mathbb{N}$, attaining target stealth measure $q_Z\in\mathcal{P}(\mathcal{Z})$, and exponent $\gamma>0$ (see Definition \ref{DEF:WTBC_achievability}). 

\noindent For every $n\in\mathbb{N}$, define $g_n\triangleq \tilde{P}^{(c_n)}_{\mathbf{X}|\mathbf{Z},M}$ and $\psi_n\triangleq\phi_n$, where $\tilde{P}^{(c_n)}_{\mathbf{X}|\mathbf{Z},M}$ is the conditional marginal distribution of $\mathbf{X}$ given $(\mathbf{Z},M)$ with respect to $\tilde{P}^{(c_n)}$ in \eqref{EQ:WTC_induced_PMF} induced by the $n$-th wiretap code $c_n$. Then:
\begin{enumerate}
    \item $b_n\triangleq (g_n,\psi_n)$ is an $(n,R)$-code for the $\big(\mathcal{Z},\mathcal{X},\mathcal{Y},q_Z,p_{Y|X,Z}\big)$ GPC.
    
    \item Letting $\tilde{Q}^{(b_n)}\triangleq\tilde{Q}^{(b_n)}_{\mathbf{Z},M,\mathbf{X},\mathbf{Y},\hat{M}}$ denote the joint distributions induced by $b_n$, for $n$ sufficiently large we have $\left\|\tilde{P}^{(c_n)}-\tilde{Q}^{(b_n)}\right\|_\mathsf{TV}\leq e^{-n\gamma}$.
    
    \item The sequence of codes $\{b_n\}_{n\in\mathbb{N}}$ attains $\mathsf{P}_\mathsf{e}(b_n)\xrightarrow[n\to\infty]{}0$, and consequently, $R$ is achievable for the aforementioned GPC.
\end{enumerate}
\end{proposition}

This proposition, whose proof is relegated to Section~\ref{SUBSEC:WTGP_codes_relations_proof}, enables the use of the GPC setup and known upper bounds on its capacity to furnish converse proofs for corresponding WTC scenarios. As a simple demonstration, we next give an alternative proof for the converse of the WTC secrecy-capacity theorem (when a stealth condition is present).

\vspace{2mm}

\subsubsection{\underline{New Converse Proof for the Classic WTC}}\label{SUBSUBSEC:PTP_Analogy_Illustration}

Consider the $\big(\mathcal{X},\mathcal{Y},\mathcal{Z},p_{Y,Z|X}\big)$ WTC, for which codes, achievability and secrecy-capacity are defined in accordance with the previous section. The secrecy-capacity $C_\mathsf{WT}(p_{Y,Z|X})$ is given by
\begin{equation}
    C_\mathsf{WT}(p_{Y,Z|X})=\max_{p_{U,X}\in\mathcal{P}(\mathcal{U}\times\mathcal{X})}\Big[I_p(U;Y)-I_p(U;Z)\Big],\label{EQ:WTC_capacity}
\end{equation}
where the mutual information terms are taken with respect to the joint distribution $p\triangleq p_{U,X}p_{Y,Z|X}$. This classic formula was first derived by Csisz{\'a}r and K{\"o}rner under the weak-secrecy measure \cite{Csiszar_Korner_BCconfidential1978}, and was shown to also hold under more stringent security notions such as strong-secrecy \cite{Bloch_Resolvability_Secrecy2013}, effective secrecy \cite{Kramer_EffectiveSecrecy2014} and even semantic-security \cite{Vardy_Semantic_WTC2012,Goldfeld_WTCII_semantic2015}. This formula also characterizes the secrecy-capacity under our definition of achievability (see \eqref{EQ:WTC_achievability_def}). The direct part is a simple consequence of the achievability proof of Theorem 1 from \cite{Kramer_EffectiveSecrecy2014} and Proposition \ref{PROP:Achievability_Relation} herein. 
In the following we show how to prove \begin{equation}
    C_\mathsf{WT}(p_{Y,Z|X})\leq\max_{p_{U,X}\in\mathcal{P}(\mathcal{U}\times\mathcal{X})}\Big[I_p(U;Y)-I_p(U;Z)\Big],\label{EQ:WTC_capacity_converse}
\end{equation}
based on having a converse proof for the GPC. Let $R$ be an achievable rate for the $\big(\mathcal{X},\mathcal{Y},\mathcal{Z},p_{Y,Z|X}\big)$ WTC and let $\{c_n\}_{n\in\mathbb{N}}$ be the corresponding sequence of $(n,R)$-codes satisfying \eqref{EQ:WTC_achievability_def} for some $\gamma>0$ and $q_Z\in\mathcal{P}(\mathcal{Z})$, and any sufficiently large $n$. By Proposition~\ref{PROP:WTGP_codes_relations}, we get a sequence of $(n,R)$-codes $\{b_n\}_{n\in\mathbb{N}}$ for the GPC $\big(\mathcal{Z},\mathcal{X},\mathcal{Y},q_Z,p_{Y|X,Z}\big)$ satisfying Items (2) and (3) from the proposition.


The converse proof for the GP coding theorem, in particular, shows that if $\{b_n\}_{n\in\mathbb{N}}$ is a sequence of $(n,R)$-codes with a vanishing error probability, then 
\begin{align*}
    R\leq \frac{1}{n}\sum_{i=1}^n&\Big[I_{\tilde{Q}^{(b_n)}}\big(M,Y_{[1:i-1]},Z_{[i+1:n]};Y_i\big)\\&-I_{\tilde{Q}^{(b_n)}}\big(M,Y_{[1:i-1]},Z_{[i+1:n]};Z_i\big)\Big]+\epsilon_n,\numberthis\label{EQ:WTC_converse_proof_UB1}
\end{align*}
where $\epsilon_n\triangleq\frac{1}{n}+R\cdot\mathsf{P}_\mathsf{e}(b_n)\xrightarrow[n\to\infty]{}0$. Let $n$ be sufficiently large so that \eqref{EQ:WTC_achievability_def} holds. Since
\begin{align*}
    \left\|\tilde{P}^{(c_n)}_{M,Y_{[1:i]},Z_{[i:n]}}-\tilde{Q}^{(b_n)}_{M,Y_{[1:i]},Z_{[i:n]}}\right\|_\mathsf{TV}&\leq\left\|\tilde{P}^{(c_n)}-\tilde{Q}^{(b_n)}\right\|_\mathsf{TV}\\&\leq e^{-n\gamma},\quad \forall i\in[1:n],\numberthis
\end{align*}
and because exponential convergence of total variation (over finite probability spaces) dominates the corresponding difference of mutual information terms (see Lemma~\ref{LEMMA:MI_continuity}), there exists a $\delta>0$, such that
\begin{align*}
    &I_{\tilde{Q}^{(b_n)}}\big(M,Y_{[1:i-1]},Z_{[i+1:n]};Y_i\big)\\&\quad\quad\quad-I_{\tilde{Q}^{(b_n)}}\big(M,Y_{[1:i-1]},Z_{[i+1:n]};Z_i\big)\\&\leq I_{\tilde{P}^{(c_n)}}\big(M,Y_{[1:i-1]},Z_{[i+1:n]};Y_i\big)\\&\quad\quad\quad-I_{\tilde{P}^{(c_n)}}\big(M,Y_{[1:i-1]},Z_{[i+1:n]};Z_i\big)+2e^{-n\delta}.\numberthis\label{EQ:WTC_converse_proof_UB2}
\end{align*}

Denoting $\eta_n\triangleq 2e^{-n\delta}+\epsilon_n$, and inserting \eqref{EQ:WTC_converse_proof_UB2} into \eqref{EQ:WTC_converse_proof_UB1}, further gives
\begin{align*}
    R&\leq \frac{1}{n}\sum_{i=1}^n\Big[I_{\tilde{P}^{(c_n)}}\big(M,Y_{[1:i-1]},Z_{[i+1:n]};Y_i\big)\\&\quad\quad\quad\quad\quad\quad-I_{\tilde{P}^{(c_n)}}\big(M,Y_{[1:i-1]},Z_{[i+1:n]};Z_i\big)\Big]+\eta_n\\
    &\stackrel{(a)}=\frac{1}{n}\sum_{i=1}^n\Big[I_{\tilde{P}^{(c_n)}}\big(U_i;Y_i\big)-I_{\tilde{P}^{(c_n)}}\big(U_i;Z_i\big)\Big]+\eta_n,\numberthis\label{EQ:WTC_converse_proof_UB3}
\end{align*}
where (a) is by defining $U_i\triangleq \left(M,Y_{[1:i-1]},Z_{[i+1:n]}\right)$, for $i\in[1:n]$. By letting $T\sim\mathsf{Unif}[1:n]$ be independent of $(M,X^n,Y^n,Z^n)$ and defining $U\triangleq(U_T,T)$, $X\triangleq X_T$, $Y\triangleq Y_T$ and $Z\triangleq Z_T$, we obtain
\begin{equation}
    R\leq I_{\tilde{p}}(U;Y)-I_{\tilde{p}}(U;Z)+\eta_n,\label{EQ:WTC_converse_proof_UB4}
\end{equation}
where $\tilde{p}$ denotes the joint distribution of the random variables $(U,X,Y,Z)$ defined above. It is readily verified that under $\tilde{p}$, $(Y,Z)- X- U$ forms a Markov chain. The converse proof is concluded by further maximizing the RHS of \eqref{EQ:WTC_converse_proof_UB4} over all $p_{U,X}\in\mathcal{P}(\mathcal{U}\times\mathcal{X})$ and taking the limit of $n\to\infty$.

\begin{remark}[Alternative Analogy-Based Derivation]
An alternative way to prove \eqref{EQ:WTC_capacity_converse} based on the analogy between the WTC and the GPC is the following. One could proceed from \eqref{EQ:WTC_converse_proof_UB1} to further upper bound $R$ as
\begin{align*}
        R&\stackrel{(a)}\leq\frac{1}{n}\sum_{i=1}^n\Big[I_{\tilde{Q}^{(b_n)}}\big(U_i;Y_i\big)-I_{\tilde{Q}^{(b_n)}}\big(U_i;Z_i\big)\Big]+\epsilon_n\\&
        \stackrel{(b)}=I_{\tilde{q}}(U;Y)-I_{\tilde{q}}(U;Z)+\epsilon_n,\numberthis\label{EQ:WTC_converse_proof_alt_UB1}
\end{align*}
where (a) is by defining $U_i\triangleq \left(M,Y_{[1:i-1]},Z_{[i+1:n]}\right)$ for $i\in[1:n]$, while (b) follows by time-sharing arguments and setting $\tilde{q}$ as the joint distribution of the random variables $U\triangleq(U_T,T)$, $X\triangleq X_T$, $Y\triangleq Y_T$ and $Z\triangleq Z_T$; here $T\sim\mathsf{Unif}[1:n]$ is the same time-sharing random variable used for the derivation of \eqref{EQ:WTC_converse_proof_UB4}. Noting that $\tilde{q}$ can be computed from $p^{(\mathsf{U})}_{[1:n]}\times\tilde{Q}^{(b_n)}$ and since the latter is well-approximated by $p^{(\mathsf{U})}_{[1:n]}\times\tilde{P}^{(c_n)}$, one can upper bound the RHS of \eqref{EQ:WTC_converse_proof_alt_UB1} with $I_{\tilde{p}}(U;Y)-I_{\tilde{p}}(U;Z)$ plus a vanishing term. Here $\tilde{p}$ is the distribution defined after \eqref{EQ:WTC_converse_proof_UB4}.

\begin{remark}[The Obtained Bound on Secrecy-Capacity]
We point out the following detail concerning our analogy-based proof. Provided an upper bound $\mathsf{U}$ on the capacity of a GPC $C_{\mathsf{GP}}(q_Z,q_{Y|X,Z})$ our method shows that the analogous WTC has $C_{\mathsf{WT}}(p_{Y,Z|X})\leq \mathsf{U}'$, for some $\mathsf{U}'\leq\mathsf{U}$. Clearly, $\mathsf{U}$ also bounds the WTC capacity, but $\mathsf{U}'$ could potentially be strictly smaller (because of the different optimizations domains). 
\end{remark}
\end{remark}

\subsection{Extension to Multiuser Setups}


We extend the above ideas to broadcast channels. By relating the WTBC and the GPBC through a claim in the spirit of Proposition~\ref{PROP:WTGP_codes_relations}, the analogy-based method enables a converse proof for any WTBC whose analogous GPBC has a solution. This is the main idea behind the proofs of Theorems~\ref{TM:WTBC_capacity} and~\ref{TM:PD_WTBC_capacity} (see Sections~\ref{SUBSEC:WTBC_capacity_proof} and~\ref{SUBSEC:PD_WTBC_capacity_proof}).

Consider a WTBC $\big(\mathcal{X},\mathcal{Y}_1,\mathcal{Y}_2,\mathcal{Z},p_{Y_1,Y_2,Z|X}\big)$ as defined in Section~\ref{SUBSEC:WTBC_def} and a corresponding sequence of stealth-wiretap codes attaining a stealth target measure $q_Z$. Similarly to the base case, an analogous GPBC is constructed in three steps:
\begin{enumerate}
\item Replace the eavesdropper of the WTBC with a state sequence $\mathbf{Z}\sim q_Z^n$;
\item Non-causally reveal $\mathbf{Z}$ to the encoder;
\item Set the GPBC's transition probability as the conditional marginal distribution $p_{Y_1,Y_2|X,Z}$. 
\end{enumerate}
The produced analogous $\big(\mathcal{Z},\mathcal{X},\mathcal{Y}_1,\mathcal{Y}_2,q_Z,p_{Y_1,Y_2|X,Z}\big)$ GPBC inherits the properties the WTBC possesses (e.g., deterministic components, order of degradeness, etc). For example, if the WTBC is SD, i.e., $p_{Y_1,Y_2,Z|X}=\mathds{1}_{\{Y_1=y_1(X)\}}p_{Y_2,Z|X}$, then so is the GPBC since $p_{Y_1,Y_2|X,Z}=\mathds{1}_{\{Y_1=y_1(X)\}}p_{Y_2|X,Z}$. If one of the observed signals of the legitimate receivers is a degraded version of the other, then the same ordering applies for the signal intercepted by the receivers of the GPBC. The analogy also accounts for WTBC settings with cooperative components. Namely, if the receivers of the WTBC are connected by, e.g., a finite-capacity noiseless communication link, then the same applies for the receivers of the analogous GPBC.

As for the base case, the WTBC secrecy-capacity regions we derive in this work correspond to those of the analogous GPBCs . Namely, the information measures bounding the transmission rates in each region have the same form. However, the underlying distributions and the optimization domains are different. These relations between the admissible rate regions are emphasized in Section~\ref{SEC:Results}.


\begin{figure*}[!t]
	\begin{center}
	    \begin{psfrags}
	        \psfragscanon
	        \psfrag{A}[][][0.9]{$(M_1,M_2)$}
	        \psfrag{J}[][][0.9]{$(M_1,M_2)$}
	        \psfrag{B}[][][1]{$f_n$}
	        \psfrag{K}[][][1]{$g_n$}
	        \psfrag{C}[][][1]{\ \ \ $\mathbf{X}$}
	        \psfrag{L}[][][1]{\ \ \ $\mathbf{X}$}
	        \psfrag{D}[][][0.8]{\ \ \ \ \ \ \ \ $p^n_{Y_1,Y_2,Z|X}$}
	        \psfrag{M}[][][0.8]{\ \ \ \ \ \ \ \ $p^n_{Y_1,Y_2|X,Z}$}
	        \psfrag{E}[][][1]{\ \ \ \ $\mathbf{Y}_1$}
	        \psfrag{N}[][][1]{\ \ \ \ $\mathbf{Y}_1$}
	        \psfrag{U}[][][1]{\ $\mathbf{Y}_2$}
	        \psfrag{X}[][][1]{\ \ \ \ $\mathbf{Y}_2$}
	        \psfrag{F}[][][1]{$\phi_1^{(n)}$}
	        \psfrag{O}[][][1]{$\psi_1^{(n)}$}
	        \psfrag{V}[][][1]{$\phi_2^{(n)}$}
	        \psfrag{Y}[][][1]{$\psi_2^{(n)}$}
	        \psfrag{G}[][][1]{\ \ $\hat{M}_1$}
	        \psfrag{P}[][][1]{\ \ $\hat{M}_1$}
	        \psfrag{W}[][][1]{\ \ $\hat{M}_2$}
	        \psfrag{Z}[][][1]{\ \ $\hat{M}_2$}
            \psfrag{H}[][][1]{\ \ \ \ \ \ \ \ \ \ \ \ \ \ \ \ \ \ \ \ \ \ \  ${\color{red!65!black}\mathbf{Z}\sim P^{(c_n)}_{\mathbf{Z}|M_1,M_2}\approx q_Z^n}$}
            \psfrag{T}[][][1]{\ \ \ ${\color{green!45!black}c_{12}}$}
	        \psfrag{R}[][][1]{\ \ \ ${\color{red!65!black}q_Z^n}$}
	        \psfrag{Q}[][][1]{${\color{red!65!black}\mathbf{Z}}$}	        
	        \subfloat[]{\includegraphics[scale = .37]{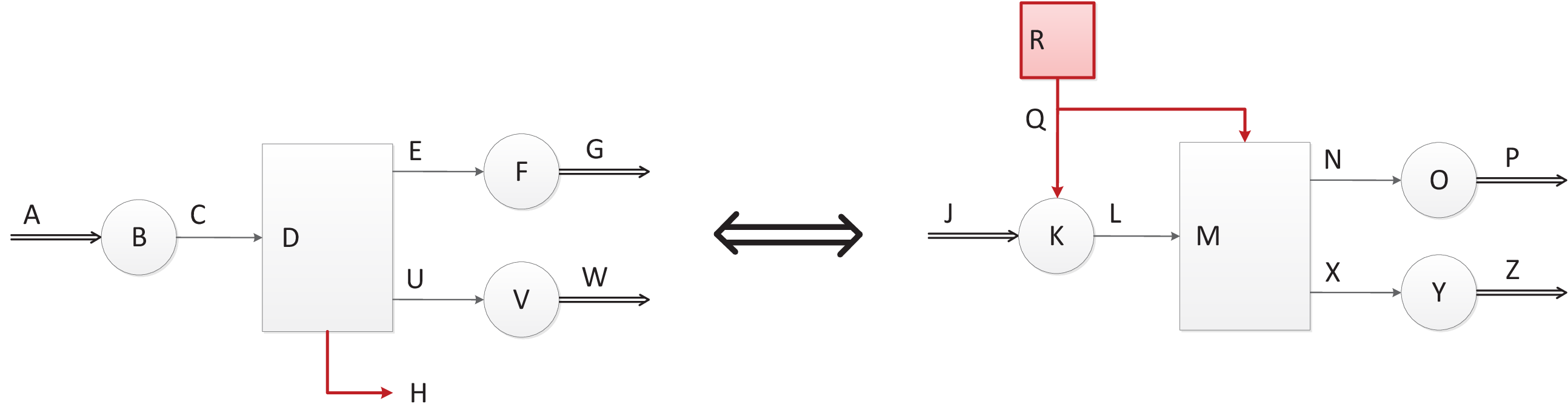}}\\
	        \subfloat[]{\includegraphics[scale = .37]{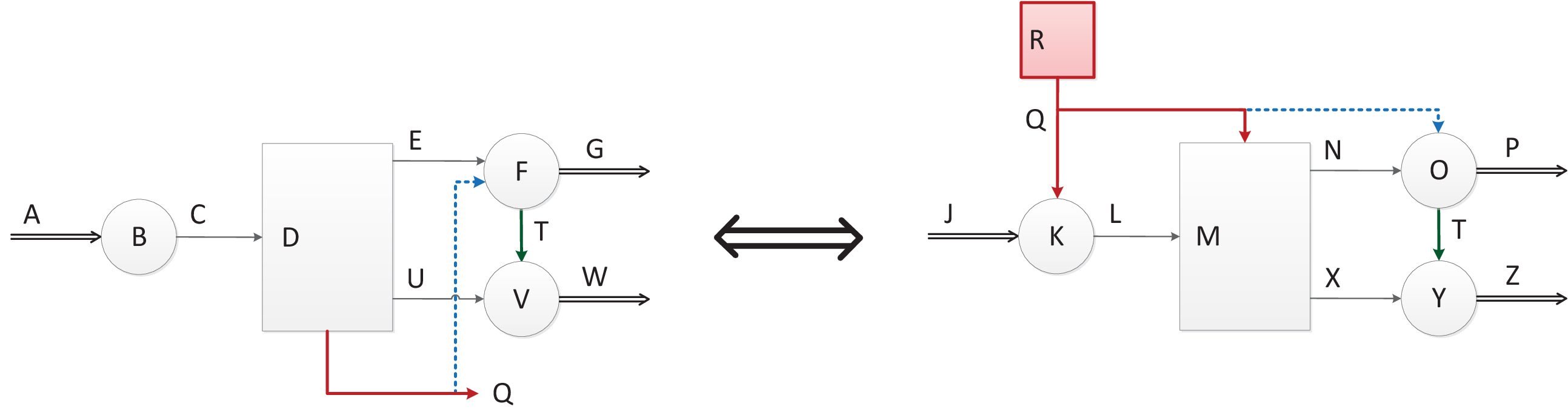}}
	        \caption{WTBC and GPBC analogy: (a) The GPBC is obtained from the WTBC by replacing the eavesdropper (whose observation distribution given the message $P^{(c_n)}_{\mathbf{Z}|M_1,M_2}$ asymptotically approaches $q_Z^n$) with a state sequence $\mathbf{Z}\sim q_Z^n$, revealing $\mathbf{Z}$ to the encoder and setting the state-dependent BC $p_{Y_1,Y_2|X,Z}$ as the conditional marginal distribution of the WTBC's transition probability $p_{Y_1,Y_2,Z|X}$; (b) If one of the WTBC's receivers also observes $\mathbf{Z}$ then the same receiver of the analogous GPBC is informed of the state. Furthermore, cooperative WTBC receivers correspond to cooperative receivers in the analogous GPBC.}\label{FIG:WT-GP-BC}
	        \psfragscanoff
	    \end{psfrags}
	\end{center}
\end{figure*}


In Fig.~\ref{FIG:WT-GP-BC}(a) we illustrate a pair of analogous WTBC and GPBC setups. The capacity region of the SD-GPBC, i.e., when $p_{Y_1,Y_2|X,Z}=\mathds{1}_{\{Y_1=y_1(X,Z)\}}p_{Y_2|X,Z}$, was derived in~\cite{Lapidoth_senideterministic2012}. Fig.~\ref{FIG:WT-GP-BC}(b) depicts the analogy when Receiver 1 of the BC is informed (of the eavesdropper's observation for the wiretap setup and of the state sequence for the GP scenario), and a cooperation link of capacity $c_{12}<\infty$ extends from Receiver 1 to Receiver 2. The capacity region of this GPBC setup without and with the cooperation link was found in~\cite{Steinberg_Degraded_BC_State2005} and~\cite{Dikstein_PDBC_Cooperation2016}, respectively, under the assumption that $Y_2-Y_1-(X,Z)$ forms a Markov chain.

Since GPBCs have been extensively treated in the literature and capacity results are available for many cases~\cite{Lapidoth_senideterministic2012,Steinberg_Degraded_BC_State2005,Dikstein_PDBC_Cooperation2016,ramachandran2017capacity}, the analogy allows leveraging these results to study corresponding WTBCs. This is done by relating the performance of two analogous models through a simple extension of Proposition~\ref{PROP:WTGP_codes_relations}, which we state next. The proof is omitted as it closely follows the derivation shown in Appendix~\ref{SUBSEC:WTGP_codes_relations_proof}.

\begin{proposition}[From WTBC Codes to GPBC Codes]\label{PROP:WTGPBC_codes_relations}
Let $\mathcal{X},\ \mathcal{Y}_1,\ \mathcal{Y}_2$ and $\mathcal{Z}$ be finite sets and consider a $\big(\mathcal{X},\mathcal{Y}_1,\mathcal{Y}_2,\mathcal{Z},p_{Y_1,Y_2,Z|X}\big)$ WTBC. Let $(R_1,R_2)\in\mathbb{R}^2_+$ be an achievable rate pair for the WTBC, with a corresponding sequence of $(n,R_1,R_2)$-codes $\{c_n\}_{n\in\mathbb{N}}$, where $c_n=\left(f_n,\phi^{(n)}_1,\phi^{(n)}_2\right)$, for each $n\in\mathbb{N}$, attaining target stealth measure $q_Z\in\mathcal{P}(\mathcal{Z})$, and exponent $\gamma>0$. For every $n\in\mathbb{N}$, define $g_n\triangleq P^{(c_n)}_{\mathbf{X}|\mathbf{Z},M_1,M_2}$ and $\psi_j^{(n)}\triangleq\phi^{(n)}_j$, for $j=1,2$, where $P^{(c_n)}_{\mathbf{X}|\mathbf{Z},M_1,M_2}$ is the conditional marginal distribution of $\mathbf{X}$ given $(\mathbf{Z},M_1,M_2)$ with respect to $P^{(c_n)}$ in \eqref{EQ:WTBC_induced_PMF} induced by the $n$-th wiretap code $c_n$. Then:
\begin{enumerate}
    \item $b_n\triangleq (g_n,\psi_1^{(n)},\psi_2^{(n)})$ is an $(n,R_1,R_2)$-code for the $\big(\mathcal{Z},\mathcal{X},\mathcal{Y}_1,\mathcal{Y}_2,q_Z,p_{Y_1,Y_2|X,Z}\big)$ GPBC.

    \item For any $n$ large enough, the distribution $Q^{(b_n)}$ induced by $b_n$ (see \eqref{EQ:GPBC_induced_PMF}) satisfies $\left\|P^{(c_n)}-Q^{(b_n)}\right\|_\mathsf{TV}\leq e^{-n\gamma}$.
    
    \item The sequence of codes $\{b_n\}_{n\in\mathbb{N}}$ attains $\mathsf{P}_\mathsf{e}(b_n)\xrightarrow[n\to\infty]{}0$, and consequently, $(R_1,R_2)$ is an achievable rate pair for the aforementioned GPBC.
\end{enumerate}
\end{proposition}


\section{New Secrecy-Capacity Results}\label{SEC:Results}

We give a single-letter characterization of the secrecy-capacity region of the SD-WTBC and of the PD-WTBC with an informed receiver (with and without receiver cooperation). In accordance to our definition of achievability, the derived regions are valid when a stealth requirement is added to the regular vanishing information leakage criterion. A discussion comparing our derivations to previously known results for these WTBC models is given in Remark \ref{REM:comparison} at the end of this section. 


\subsubsection{\underline{Semi-Deterministic WTBCs}}\label{SUBSUBSEC:SD_WTBC_results}

Consider an $\big(\mathcal{X},\mathcal{Y}_1,\mathcal{Y}_2,\mathcal{Z},\mathds{1}_{\{Y_1=y_1(X)\}}p_{Y_2,Z|X}\big)$ SD-WTBC and let $\mathcal{U}$ be a set with cardinality $|\mathcal{U}|\leq|\mathcal{X}|+1$. For any $p_{U,X}\in\mathcal{P}(\mathcal{U}\times\mathcal{X})$ define
\begin{align*}
&\mathcal{R}^{(\mathsf{SD})}_\mathsf{WT}\left(p_{U,X},\mathds{1}_{\{Y_1=y_1(X)\}}p_{Y_2,Z|X}\right)\\
&\triangleq\left\{ (R_1,R_2)\in\mathbb{R}_+^2 \mspace{3mu}\Vast|\mspace{3mu}\begin{aligned}
    R_1 &\leq H_p(Y_1|Z), \\
    R_2 &\leq I_p(U;Y_2)-I_p(U;Z), \\
    R_1+R_2 &\leq H_p(Y_1|Z)+I_p(U;Y_2)\\&\quad\quad\quad\quad\mspace{6mu}-I_p(U;Y_1,Z)
\end{aligned}
\right\},\numberthis\label{EQ:WTBC_region_prob}
\end{align*}
where the mutual information terms are calculated with respect to the joint distribution $p\triangleq p_{U,X}\mathds{1}_{\{Y_1=y_1(X)\}}p_{Y_2,Z|X}$, under which $(Y_1,Y_2,Z)- X- U$ forms a Markov chain and $Y_1$ is a deterministic function of $X$.

\begin{theorem}[SD-WTBC Secrecy-Capacity]\label{TM:WTBC_capacity}  
The secrecy-capacity region of the $\big(\mathcal{X},\mathcal{Y}_1,\mathcal{Y}_2,\mathcal{Z},\mathds{1}_{\{Y_1=y_1(X)\}}$ $\times p_{Y_2,Z|X}\big)$ SD-WTBC is given by
\begin{align*}
    &\mathcal{C}^{(\mathsf{SD})}_\mathsf{WT}\left(\mathds{1}_{\{Y_1=y_1(X)\}}p_{Y_2,Z|X}\right)\\&=\bigcup_{p_{U,X}\in\mathcal{P}(\mathcal{U}\times\mathcal{X})}\mathcal{R}^{(\mathsf{SD})}_\mathsf{WT}\left(p_{U,X},\mathds{1}_{\{Y_1=y_1(X)\}}p_{Y_2,Z|X}\right).\numberthis\label{EQ:WTBC_capacity}
\end{align*}
\end{theorem}

The proof of Theorem~\ref{TM:WTBC_capacity} is given in Section~\ref{SUBSEC:WTBC_capacity_proof}. The direct part relies on a specialization of the inner bound on the secrecy-capacity region of the WTBC derived in~\cite[Theorem 3]{Yassaee_random_binning2014}. As the performance criterion in that work corresponds to the definition of achievability used herein (Definition~\ref{DEF:WTBC_achievability}), the result from~\cite{Yassaee_random_binning2014} applies for our setup.\footnote{While the definition of WTBC achievability in \cite{Yassaee_random_binning2014} does not explicitly require stealth (see Section IV-F therein), their proof attains it nonetheless. To see this, note that the $p(z^n)$ used throughout the proof of Theorem 3 from \cite{Yassaee_random_binning2014} is a product distribution (see Protocol A) and that their achievability proof in particular shows that the distribution induced by the code at the eavesdropper's end approximates this $p(z^n)$ (as evident from Equation (52) in \cite{Yassaee_random_binning2014} and the two unnumbered equations following it).} The main technical novelty in this work is the proof of the converse claim. Namely, we relate the WTBC to the corresponding GPBC that was studied in~\cite{Lapidoth_senideterministic2012} and use the converse proof of Theorem 1 therein to establish the RHS of \eqref{EQ:WTBC_capacity} as an outer bound on $\mathcal{C}^{(\mathsf{SD})}_\mathsf{WT}\left(\mathds{1}_{\{Y_1=y_1(X)\}}p_{Y_2,Z|X}\right)$.

\begin{remark}[Analogous SD-GPBC] The analogous GPBC is defined by the tuple $\big(\mathcal{Z},\mathcal{X},\mathcal{Y}_1,\mathcal{Y}_2,q_Z,\mathds{1}_{\{Y_1=y_1(X)\}}q_{Y_2|X,Z}\big)$. It was considered in~\cite{Lapidoth_senideterministic2012}, where its capacity region was found. The region is described by the same inequalities as in \eqref{EQ:WTBC_region_prob}, but with an underlying distribution $q_Zq_{U,X|Z}\mathds{1}_{\{Y_1=y_1(X)\}}q_{Y_2|X,Z}$, and the union is over all $q_{U,X|Z}$. Evidently, the structure of the two regions correspond.

We note that the converse proof of Theorem 1 from \cite{Lapidoth_senideterministic2012} is highly nontrivial and relies on novel optimization tricks for arriving at an outer bound described by a single auxiliary variable. Furthermore, the proof makes use of the i.i.d. nature of the state sequence $\mathbf{Z}$ and its independence of the messages multiple times. Through the analogy, we are able to `borrow' these nice statistical properties of $\mathbf{Z}$ and the messages, as well as the novel optimization tricks to derive the WTBC converse of Theorem \ref{TM:WTBC_capacity} above. We stress that one may not simply repeat the GPBC converse proof steps when upper bounding the transmission rates over a WTBC. This is since the underlying distribution in the WTBC is $P^{(c_n)}$ from \eqref{EQ:WTBC_induced_PMF}, under which $\mathbf{Z}$ it is generally not i.i.d. nor it is independent of $(M_1,M_2)$.


\end{remark}

\begin{remark}[Cardinality Bound]\label{REM:SD_WTBC_cardinality}
Theorem~\ref{TM:WTBC_capacity} implicitly states that to exhaust the secrecy-capacity region one may restrict the auxiliary random variable $U$ to take values in a set $\mathcal{U}$ whose cardinality is bounded by $|\mathcal{U}|\leq |\mathcal{X}|+1$. This is a direct consequence of the
Eggleston-Fenchel-Carath{\'e}odory theorem~\cite[Theorem 18]{Eggleston_Convexity1958} via an approach identical to the applied for the analogous GPBC  in~\cite[Appendix A]{Lapidoth_senideterministic2012conf}.

\end{remark}


\subsubsection{\underline{Physically Degraded WTBCs with an Informed Receiver}}

Consider an $\big(\mathcal{X},\mathcal{Y}_1,\mathcal{Y}_2,\mathcal{Z},p_{Y_1,Z|X}p_{Y_2|Y_1}\big)$ PD-WTBC with an informed receiver, i.e., when Receiver 1 observes the pair $(Y_1,Z)$. Let $\mathcal{U}$ be a set with cardinality $|\mathcal{U}|\leq|\mathcal{X}|+1$, and for any $p_{U,X}\in\mathcal{P}(\mathcal{U}\times\mathcal{X})$ define
\begin{align*}
&\mathcal{R}^{(\mathsf{PD-IR})}_\mathsf{WT}\left(p_{U,X},p_{Y_1,Z|X}p_{Y_2|Y_1}\right)\\&\mspace{5mu}\triangleq\left\{ (R_1,R_2)\in\mathbb{R}_+^2 \mspace{3mu}\vasti|\mspace{3mu}\begin{aligned}
    R_1 &\leq I_p(X;Y_1|U,Z), \\
    R_2 &\leq I_p(U;Y_2)-I_p(U;Z)
\end{aligned}
\right\},\numberthis\label{EQ:PD_WTBC_region_prob}
\end{align*}
where the mutual information terms are calculated with respect to the joint distribution $p\triangleq p_{U,X}p_{Y_1,Z|X}p_{Y_2|Y_1}$, under which $Y_2 - Y_1 - (Z,X,U)$ and $(Y_1,Y_2,Z) - X - U$ form Markov chains.

\begin{theorem}[PD-WTBC Secrecy-Capacity]\label{TM:PD_WTBC_capacity}  
The secrecy-capacity region of the $\big(\mathcal{X},\mathcal{Y}_1,\mathcal{Y}_2,\mathcal{Z},p_{Y_1,Z|X}p_{Y_2|Y_1}\big)$ PD-WTBC with an informed receiver is given by
\begin{align*}
    &\mathcal{C}^{(\mathsf{PD-IR})}_\mathsf{WT}\left(p_{Y_1,Z|X}p_{Y_2|Y_1}\right)\\&\quad\quad=\bigcup_{p_{U,X}\in\mathcal{P}(\mathcal{U}\times\mathcal{X})}\mathcal{R}^{(\mathsf{PD-IR})}_\mathsf{WT}\left(p_{U,X},p_{Y_1,Z|X}p_{Y_2|Y_1}\right).\numberthis\label{EQ:PD_WTBC_capacity}
\end{align*}
\end{theorem}

Theorem~\ref{TM:PD_WTBC_capacity} is proven in Section~\ref{SUBSEC:PD_WTBC_capacity_proof}. The converse is derived via analogy-based arguments that exploit the capacity characterization for the corresponding PD-GPBC with an informed decoder~\cite[Theorem 3]{Steinberg_Degraded_BC_State2005} (see the following remark). Achievability follows by random coding arguments that use a stealth-wiretap superposition code construction. The inner layer encodes the $M_2$ message and is decoded by both receivers. The message $M_1$ is encoded in the outer layer of the code and is decoded by the informed receiver only. Both layers of the code are injected with sufficient randomness to conceal the confidential messages from the eavesdropper.

Although achievability of $\mathcal{C}^{(\mathsf{PD-IR})}_\mathsf{WT}\left(p_{Y_1,Z|X}p_{Y_2|Y_1}\right)$ can be derived as a consequence of~\cite[Theorem 3]{Yassaee_random_binning2014}, we prefer to prove it directly through the aforementioned superposition coding scheme. The explicit construction readily extends to account for cooperative receivers (through a simple binning argument). This extension enables deriving achievability for the cooperative version of the considered PD-WTBC, as explained in the next subsection.

\begin{remark}[Analogous PD-GPBC] The analogous $\big(\mathcal{Z},\mathcal{X},\mathcal{Y}_1,\mathcal{Y}_2,q_Z,q_{Y_1|X,Z}q_{Y_2|Y_1}\big)$ PD-GPBC with an informed receiver was studied in~\cite[Theorem 3]{Steinberg_Degraded_BC_State2005}. Specifically, non-casual CSI was assumed to be available at the encoder and at Receiver 1. The capacity region is described by the same expressions as in \eqref{EQ:PD_WTBC_region_prob}, but with an underlying distribution $q\triangleq q_Zq_{U,X|Z}q_{Y_1|X,Z}q_{Y_2|Y_1}$ with the union taken over all $q_{U,X|Z}$ distributions. 
\end{remark}

\begin{remark}[Cardinality Bound]
The cardinality bound on $\mathcal{U}$ again follows from the
Eggleston-Fenchel-Carath{\'e}odory theorem~\cite[Theorem 18]{Eggleston_Convexity1958} through arguments similar to those presented in Remark~\ref{REM:SD_WTBC_cardinality}. The proof is omitted.
\end{remark}

\subsubsection{\underline{PD-WTBC with Informed Receiver and Cooperation}}

A simple extension of the proof of Theorem~\ref{TM:PD_WTBC_capacity} produces a single-letter characterization of the secrecy-capacity of the considered PD-WTBC with cooperative receivers. More specifically, consider the PD-WTBC with an informed receiver and where a unidirectional noiseless link of capacity $c_{12}<\infty$ extends from (the informed) Receiver 1 to (the uninformed) Receiver 2. 

\begin{corollary}[Cooperative PD-WTBC Secrecy-Capacity]\label{CORR:PD_WTBC_capacity_coop} The secrecy-capacity region of this setup is the union of rate pairs $(R_1,R_2)\in\mathbb{R}^2_+$ satisfying:
\begin{subequations}
\begin{align}
    R_1 &\leq I_p(X;Y_1|U,Z)\label{EQ:PD_WTBC_coop_RB1} \\
    R_2 &\leq I_p(U;Y_2)-I_p(U;Z)+c_{12}\label{EQ:PD_WTBC_coop_RB2}\\
    R_1+R_2 &\leq I_p(X;Y_1|Z)\label{EQ:PD_WTBC_coop_RB12}
\end{align}\label{EQ:PD_WTBC_coop_RB}%
\end{subequations}
where the union is over all $p_{U,X}\in\mathcal{P}(\mathcal{U}\times\mathcal{X})$, each inducing a joint distribution $p\triangleq p_{U,X}p_{Y_1,Z|X}p_{Y_2|Y_1}$. 
\end{corollary}

A derivation of this result is omitted due to its similarity to the proof of Theorem~\ref{TM:PD_WTBC_capacity}. Roughly speaking, achievability follows by the same stealth-wiretap superposition code used for the non-cooperative case, with one additional ingredient. To exploit the cooperation link, the legitimate parties agree on a partitioning of the message set $\mathcal{M}_2^{(n)}$ into approximately $2^{nc_{12}}$ equal sized subsets (bins). Upon decoding the inner layer of the code (which carries $M_2$), Receiver 1 shares the bin index of the decoded $M_2$ with Receiver 2 via the cooperation link. This reduces the search space of Receiver 2 in decoding $M_2$ by a factor of $2^{-nc_{12}}$, and produces  \eqref{EQ:PD_WTBC_coop_RB2}. With this modification, the reliability and security analysis are similar to those presented in Section~\ref{SUBSEC:PD_WTBC_capacity_proof}. For the converse, we use the analogy to the PD-GPBC with cooperative receivers discussed in the next remark.

\begin{remark}[Analogous PD-GPBC]\label{REM:PD_GPBC_coop} An instance of the considered PD-GPBC where the receivers can cooperate via a unidirectional finite-capacity noiseless link that extends from (the informed) Receiver 1 to (the uninformed) Receiver 2 was studied in~\cite{Dikstein_PDBC_Cooperation2016}. Theorem 7 from that work characterizes the admissible rate region by the same expressions as in \eqref{EQ:PD_WTBC_coop_RB}, but with an underlying distribution $q\triangleq q_Zq_{U,X|Z}q_{Y_1|X,Z}q_{Y_2|Y_1}$. 
\end{remark}

\begin{remark}[Comparison to Past WTBC Results]\label{REM:comparison}
The secrecy-capacity regions of the SD- and the PD-WTBCs are currently known only in some special cases. Notably,~\cite[Theorem 5]{Piantanida_External_Eve2015} characterized the SD-WTBC's weak-secrecy-capacity region under the assumption that the stochastic channel to the second receiver is less noisy than the eavesdropper's channel. Theorem 7 in that paper established the optimal region of the (non-cooperative) PD-WTBCs when both legitimate users are less noisy than the eavesdropper. 

Because our notion of security includes also a stealth condition, the secrecy-capacity regions derived herein are generally inner bounds on the corresponding regions from \cite{Piantanida_External_Eve2015}. Nonetheless, it is noteworthy that our result for the SD-WTBC does not rely on the existence of any ordering between the channels to the different nodes. Similarly, for the PD-WTBC with an informed receiver our derivation avoids the assumption that the uninformed receiver observes a better signal than the eavesdropper. Furthermore, our secrecy-capacity descriptions use a single auxiliary random variable, as opposed to the two auxiliaries used in Theorems 5 and 7 of~\cite{Piantanida_External_Eve2015}. The converse proofs of these two theorems directly upper bound the transmission rates under the WTBC's distribution; getting tight bounds relies on the less-noisy assumptions and the second auxiliary. By virtue of our analogy, we avoid directly upper bounding the WTBC's transmission rates, and instead, exploit the GPBC's converse proof from~\cite[Theorem 1]{Lapidoth_senideterministic2012} and~\cite[Theorem 3]{Dikstein_PDBC_Cooperation2016} to establish the converses of our Theorems~\ref{TM:WTBC_capacity} and~\ref{TM:PD_WTBC_capacity}, respectively. Consequently, our converse proofs use only a single auxiliary (as in~\cite[Theorem 1]{Lapidoth_senideterministic2012} and \cite[Theorem 3]{Steinberg_Degraded_BC_State2005}) and do not assume any ordering between the sub-channels beyond what is required in the analogous GP scenarios.

\end{remark}

\section{Proofs}\label{SEC:proofs}

\subsection{Proposition~\ref{PROP:WTGP_codes_relations}}\label{SUBSEC:WTGP_codes_relations_proof}

To prove (1) simply note that for each $n\in\mathbb{N}$, $\tilde{P}^{(c_n)}_{\mathbf{X}|\mathbf{Z},M}:\mathcal{Z}^n\times\mathcal{M}_n\to\mathcal{P}(\mathcal{X}^n)$ and $\phi_n:\mathcal{Y}^n\to\mathcal{M}_n$. Therefore, $g_n$ and $\psi_n$ are valid (stochastic) encoding and (deterministic) decoding functions for the GPC.

For (2), fix $n\in\mathbb{N}$, and first observe
\begin{align*}
    \tilde{P}^{(c_n)}_{M,\mathbf{X},\mathbf{Y},\mathbf{Z},\hat{M}}\mspace{-1mu}&=\mspace{-1mu}\tilde{P}^{(c_n)}_{M,\mathbf{Z}}\cdot\tilde{P}^{(c_n)}_{\mathbf{X},\mathbf{Y},\hat{M}|M,\mathbf{Z}}\\
    &\stackrel{(a)}=\mspace{-1mu}\tilde{P}^{(c_n)}_{M,\mathbf{Z}}\cdot g_n\cdot p^n_{Y|X,Z}\cdot \mathds{1}_{\big\{\hat{M}=\psi_n(\mathbf{Y})\big\}}\\
    &\stackrel{(b)}=\mspace{-1mu}\tilde{P}^{(c_n)}_{M,\mathbf{Z}}\cdot \tilde{Q}^{(b_n)}_{\mathbf{X},\mathbf{Y},\hat{M}|M,\mathbf{Z}}
\end{align*}
where (a) follows by the factorization of $\tilde{P}^{(c_n)}$ in \eqref{EQ:WTC_induced_PMF}, while (b) is because $b_n=(g_n,\psi_n)$ and due to the structure of $\tilde{Q}^{(b_n)}$ in \eqref{EQ:GPC_induced_PMF}. Recalling that $\tilde{Q}^{(c_n)}_{\mathbf{Z},M}=q_Z^n\cdot p^{(\mathsf{U})}_{\mathcal{M}_n}$, we have
\begin{equation}
    \left\|\tilde{P}^{(c_n)}-\tilde{Q}^{(b_n)}\right\|_\mathsf{TV}=\left\|\tilde{P}^{(c_n)}_{M,\mathbf{Z}}-p^{(\mathsf{U})}_{\mathcal{M}_n}\cdot q_Z^n\right\|_\mathsf{TV}\xrightarrow[n\to\infty]{}0,\label{EQ:WTC_GPC_induced_close}
\end{equation}
where (a) uses Property (3-b) from Lemma~\ref{LEMMA:TV_properties} and (b) follows by the hypothesis of Proposition~\ref{PROP:WTGP_codes_relations}.

The third claim follows by noting that
\begin{align*}
&\mathsf{P}_\mathsf{e}(b_n)\\
&\stackrel{(a)}=\left\|\tilde{Q}^{(b_n)}_{M,\hat{M}}-p^{(\mathsf{U})}_{\mathcal{M}_n}\mathds{1}_{\{\hat{M}=M\}}\right\|_\mathsf{TV}\\
&\stackrel{(b)}\leq\left\|\tilde{Q}^{(b_n)}_{M,\hat{M}}-\tilde{P}^{(c_n)}_{M,\hat{M}}\right\|_\mathsf{TV}+\left\|\tilde{P}^{(c_n)}_{M,\hat{M}}-p^{(\mathsf{U})}_{\mathcal{M}_n}\mathds{1}_{\{\hat{M}=M\}}\right\|_\mathsf{TV}\\
&\stackrel{(c)}\leq\left\|\tilde{Q}^{(b_n)}-\tilde{P}^{(c_n)}\right\|_\mathsf{TV}+\left\|\tilde{P}^{(c_n)}_{M,\hat{M},\mathbf{Z}}-p^{(\mathsf{U})}_{\mathcal{M}_n}\mathds{1}_{\{\hat{M}=M\}}q_Z^n\right\|_\mathsf{TV}
\end{align*}
where (a) is by \eqref{EQ:GPC_Relation_TV_Pe}, (b) is the triangle inequality, while (c) uses Property (3-a) from Lemma~\ref{LEMMA:TV_properties}, \eqref{EQ:WTC_GPC_induced_close} and the hypothesis of Proposition~\ref{PROP:WTGP_codes_relations}.


\subsection{Proof of Theorem~\ref{TM:WTBC_capacity}}\label{SUBSEC:WTBC_capacity_proof}


\subsubsection{\underline{Direct}}

The achievability of $\mathcal{C}_\mathsf{WT}(\mathds{1}_{\{Y_1=y_1(X)\}}p_{Y_2,Z|X})$ is a consequence of Theorem 3 from~\cite{Yassaee_random_binning2014}, where an inner bound on the secrecy-capacity region of any DM-WTBC with a common message and two private messages was derived. We restate~\cite[Theorem 3]{Yassaee_random_binning2014} in the following without the common message ($R_0=0$). 

Consider the $\big(\mathcal{X},\mathcal{Y}_1,\mathcal{Y}_2,\mathcal{Z},p_{Y_1,Y_2,Z|X}\big)$ WTBC. Let $\mathcal{Q}$ and $\mathcal{U}_j$, for $j=0,1,2$, be finite sets and for any $p_{Q,U_{[0:2]},X}\in\mathcal{P}(\mathcal{Q}\times\mathcal{U}_0\times\mathcal{U}_1\times\mathcal{U}_2\times\mathcal{X})$ let $\mathcal{R}^{(\mathsf{I})}_\mathsf{WT}\left(p_{Q,U_{[0:2]},X},p_{Y_1,Y_2,Z|X}\right)$ be the set of $(R_1,R_2)\in\mathbb{R}_+^2$ satisfying
\begin{subequations}
\begin{align}
    R_1 &\leq I_p(U_0,U_1;Y_1|Q)-I_p(U_0,U_1;Z|Q)\\
    R_2 &\leq I_p(U_0,U_2;Y_2|Q)-I_p(U_0,U_2;Z|Q)\\
    R_1+R_2 &\leq \min\big\{I_p(U_0;Y_1|Q),I_p(U_0;Y_2|Q)\big\}\nonumber\\&\quad+I_p(U_1;Y_1|U_0,Q)+I_p(U_2;Y_2|U_0,Q)\nonumber\\&\quad\quad-I_p(U_1;U_2|U_0,Q)-I_p(U_{[0:2]};Z|Q)\\
    R_1+R_2 &\leq I_p(U_0,U_1;Y_1|Q)-I_p(U_0,U_1;Z|Q)\nonumber\\&\quad+I_p(U_0,U_2;Y_2|Q)-I_p(U_0,U_2;Z|Q)\nonumber\\&\quad\quad-I_p(U_1;U_2|U_0,Z,Q)
\end{align}\label{EQ:WTBC_achievable_prob}%
\end{subequations}
where the mutual information terms are taken with respect to the joint distribution $p\triangleq p_{Q,U_{[0:2]},X}p_{Y_1,Y_2,Z|X}$, under which $(Y_1,Y_2,Z)- X- \big(Q,U_{[0:2]}\big)$ forms a Markov chain. An inner bound on the secrecy-capacity region $\mathcal{C}_\mathsf{WT}(p_{Y_1,Y_2,Z|X})$ of the WTBC (with respect to the notion of achievability from Definition~\ref{DEF:WTBC_achievability}\footnote{As explained in footnote 5 of Section \ref{SUBSUBSEC:SD_WTBC_results}, although the definition of WTBC achievability in \cite{Yassaee_random_binning2014} does explicitly requires stealth (while Definition \ref{DEF:WTBC_achievability} does), their proof establishes it nonetheless.}) is 
\begin{equation}
    \mathcal{C}_\mathsf{WT}(p_{Y_1,Y_2,Z|X})\supseteq\bigcup_{p_{Q,U_{[0:2]},X}}\mathcal{R}^{(\mathsf{I})}_\mathsf{WT}\left(p_{Q,U_{[0:2]},X},p_{Y_1,Y_2,Z|X}\right).
\end{equation}
For the considered SD-WTBC $p_{Y_1,Y_2,Z|X}=\mathds{1}_{\{Y_1=y_1(X)\}}p_{Y_2,Z|X}$, setting $Q=U_0=0$, $U_1=Y_1$ and recasting $U_2$ as $U$ reduces \eqref{EQ:WTBC_achievable_prob} to \eqref{EQ:WTBC_region_prob}. Since $Y_1=y_1(X)$, this choice of the auxiliary random variables $\big(Q,U_{[0:2]}\big)$ is feasible.


\subsubsection{\underline{Converse}}\label{SUBSUBSEC:WTBC_converse_proof}

The converse proof closely follows the arguments from Section~\ref{SUBSUBSEC:PTP_Analogy_Illustration} for the classic WTC. Let $(R_1,R_2)\in\mathbb{R}_+^2$ be an achievable rate pair for the SD-WTBC and  $\{c_n\}_{n\in\mathbb{N}}$ be a sequence of $(n,R_1,R_2)$-codes satisfying \eqref{EQ:WTBC_achievability_def} for some $\gamma>0$ and $q_Z\in\mathcal{P}(\mathcal{Z})$, and any $n$ large enough. By Proposition~\ref{PROP:WTGPBC_codes_relations}, $\{c_n\}_{n\in\mathbb{N}}$ gives rise to a sequence of $(n,R_1,R_2)$-codes $\{b_n\}_{n\in\mathbb{N}}$ for the $\big(\mathcal{Z},\mathcal{X},\mathcal{Y}_1,\mathcal{Y}_2,q_Z,p_{Y_1,Y_2|X,Z}\big)$ GPBC, each inducing a joint distribution $Q^{(b_n)}$ as given in \eqref{EQ:GPBC_induced_PMF}, such that:
\begin{enumerate}
    \item $\left\|P^{(c_n)}-Q^{(b_n)}\right\|_\mathsf{TV}\leq e^{-n\gamma}$, for any sufficiently large $n$, where $P^{(c_n)}$ is the distribution in \eqref{EQ:WTBC_induced_PMF}.
    \item $\mathsf{P}_\mathsf{e}(b_n)\xrightarrow[n\to\infty]{}0$.
\end{enumerate}
Furthermore, note that since the WTBC is SD, i.e., its transition probability factors as $p_{Y_1,Y_2,Z|X}=\mathds{1}_{\{Y_1=y_1(X)\}}p_{Y_2,Z|X}$, the obtained GPBC is also SD. Namely, we have
that the GPBC's transition probability decomposes as $p_{Y_1,Y_2|X,Z}=\mathds{1}_{\{Y_1=y_1(X)\}}p_{Y_2|X,Z}$, which falls under the framework of~\cite[Theorem 1]{Lapidoth_senideterministic2012}.\footnote{The deterministic component $Y_1$ of a SD-GPBC is, generally, a function of the input-state pair $(X,Z)$. In our case, $Y_1=y_1(X)$ depends only on the input.}

The converse proof of~\cite[Theorem 1]{Lapidoth_senideterministic2012} for the SD-GPBC shows that under $Q^{(b_n)}$, we have 
\begin{subequations}
\begin{align}
    R_1&\leq \frac{1}{n}\sum_{i=1}^nH_{Q^{(b_n)}}(Y_{1,i}|Z_i)+\epsilon_n\label{EQ:WTBC_converse_proof_UB1}\\
    R_2&\leq \frac{1}{n}\sum_{i=1}^n\Big[I_{Q^{(b_n)}}\big(M_2,Y_{2,[1:i-1]},Z_{[i+1:n]};Y_{2,i}\big)\nonumber\\&-I_{Q^{(b_n)}}\big(M_2,Y_{2,[1:i-1]},Z_{[i+1:n]};Z_i\big)\Big]\mspace{-4mu}+\mspace{-3mu}\epsilon_n\label{EQ:WTBC_converse_proof_UB2}\\
    R_1\mspace{-4mu}+\mspace{-3mu}R_2&\leq \frac{1}{n}\sum_{i=1}^n\Big[H_{Q^{(b_n)}}(Y_{1,i}|Z_i)\nonumber\\
    &-I_{Q^{(b_n)}}\mspace{-2mu}\big(\mspace{-1mu}M_2,\mspace{-2mu}Y_{2,[1:i-1]},\mspace{-2mu}Z_{[i+1:n]},\mspace{-2mu}Y_{1,[i+1:n]};Z_i,Y_{2,i}\big)\nonumber\\&+I_{Q^{(b_n)}}\mspace{-2mu}\big(\mspace{-1mu}M_2,\mspace{-2mu}Y_{2,[1:i-1]},\mspace{-2mu}Z_{[i+1:n]},\mspace{-2mu}Y_{1,[i+1:n]};Y_{2,i}\big)\Big]\mspace{-4mu}+\mspace{-3mu}\epsilon_n\label{EQ:WTBC_converse_proof_UB12}
\end{align}\label{EQ:WTBC_converse_proof_UB}%
\end{subequations}
where $\epsilon_n\triangleq\frac{2}{n}+\sum_{j=1,2}R_j\cdot\mathsf{P}_\mathsf{e}(b_n)\xrightarrow[n\to\infty]{}0$. On account of Property (3-a) of Lemma~\ref{LEMMA:TV_properties}, for $n$ large we have
\begin{align*}
    \left\|P^{(c_n)}_{M,Y_{[1:i]},Z_{[i:n]}}-Q^{(b_n)}_{M,Y_{[1:i]},Z_{[i:n]}}\right\|_\mathsf{TV}&\leq\left\|P^{(c_n)}-Q^{(b_n)}\right\|_\mathsf{TV}\\&\leq e^{-n\gamma},\numberthis
\end{align*}
uniformly in $i\in[1:n]$. Combining this with \eqref{EQ:MI_continuity_implication} from Lemma~\ref{LEMMA:MI_continuity}, we may replace the information measures from the RHS of \eqref{EQ:WTBC_converse_proof_UB} that are taken with respect to $Q^{(b_n)}$ with the same terms, but with an underlying distribution $P^{(c_n)}$ plus a vanishing term. Namely, there exists a $\delta>0$, such that for $n$ large enough
\begin{subequations}
\begin{align}
    R_1&\leq \frac{1}{n}\sum_{i=1}^nH_{P^{(c_n)}}(Y_{1,i}|Z_i)+\epsilon_n+e^{-n\delta}\label{EQ:P_WTBC_converse_proof_UB1}\\
    R_2&\leq \frac{1}{n}\sum_{i=1}^n\mspace{-2mu}\Big[I_{P^{(c_n)}}(V_i;Y_{2,i})-I_{P^{(c_n)}}(V_i;Z_i)\Big]\nonumber\\&\quad\quad\quad\quad\quad\quad\quad\quad\quad\quad\quad\quad\mspace{5mu}+\epsilon_n+2e^{-n\delta}\label{EQ:P_WTBC_converse_proof_UB2}\\
    R_1+R_2&\leq \frac{1}{n}\sum_{i=1}^n\Big[H_{P^{(c_n)}}(Y_{1,i}|Z_i)+I_{P^{(c_n)}}(V_i,T_i;Y_{2,i})\nonumber\\&\quad\mspace{8mu}-I_{P^{(c_n)}}(V_i,T_i;Y_{1,i},Z_i)\Big]+\epsilon_n+3e^{-n\delta}\label{EQ:P_WTBC_converse_proof_UB12}
\end{align}\label{EQ:P_WTBC_converse_proof_UB}%
\end{subequations}
where, for every $i\in[1:n]$, we have defined $V_i\triangleq \big(M_2,Y_{2,[1:i-1]},Z_{[i+1:n]}\big)_{P^{(c_n)}}$ and $T_i\triangleq \big(Y_{1,[i+1:n]}\big)_{P^{(c_n)}}$, with the subscript $P^{(c_n)}$ indicating that the underlying distribution is the one from \eqref{EQ:WTBC_induced_PMF}. 

Letting $n$ tend to infinity in \eqref{EQ:P_WTBC_converse_proof_UB}, we see that any achievable pair $(R_1,R_2)$ is contained in the convex closure of the union of rate pairs satisfying
\begin{subequations}
\begin{align}
    R_1&\leq H_p(Y_1|Z)\label{EQ:SL_WTBC_converse_proof_UB1}\\
    R_2&\leq I_p(V;Y_2)-I_p(V;Z)\label{EQ:SL_WTBC_converse_proof_UB2}\\
    R_1+R_2&\leq H_p(Y_1|Z)+I_p(V,T;Y_2)-I_p(V,T;Y_1,Z),\label{EQ:SL_WTBC_converse_proof_UB12}
\end{align}\label{EQ:SL_WTBC_converse_proof_UB}%
\end{subequations}
where the union is over all $p_{V,T,X}\in\mathcal{P}(\mathcal{V}\times\mathcal{T}\times\mathcal{X})$, each inducing a joint distribution $p\triangleq p_{V,T,X}p_{Y_1,Y_2,Z|X}$, i.e., $(Y_1,Y_2,Z)- X- (V,T)$ forms a Markov chain. The validity of this Markov relation follows by observing that under $P^{(c_n)}$, we have
$(Y_{1,i},Y_{2,i},Z_i)- X_i - \big(M_2,Y_{1,[i+1:n]},Y_{2,[1:i-1]},Z_{[i+1:n]}\big)$, for every $i\in[1:n]$.

To prove the converse part of Theorem~\ref{TM:WTBC_capacity}, it remains to show that there exists a (single) auxiliary random variable $U$, such that for any $(V,T)$ it satisfies both:
\begin{subequations}
\begin{align}
&I_p(V;Y_2)-I_p(V;Z)\leq I_p(U;Y_2)-I_p(U;Z)\label{EQ:WTBC_converse_proof_cond1}\\
&H_p(Y_1|Z)+I_p(V,T;Y_2)-I_p(V,T;Y_1,Z)\nonumber\\
&\quad\quad\quad\quad\leq H_p(Y_1|Z)+I_p(U;Y_2)-I_p(U;Y_1,Z).\label{EQ:WTBC_converse_proof_cond2}
\end{align}
\end{subequations}

Inspired by~\cite{Lapidoth_senideterministic2012}, we show that either $U=V$ or $U=(V,T)$ will do. Assume first that $p$ is such that
\begin{equation}
I_p(T;Y_2|V)-I_p(T;Z|V)\leq 0,\label{EQ:WTBC_converse_proof_case11}
\end{equation}
which implies 
$I_p(T;Y_2|V)-I_p(T;Y_1,Z|V)\leq 0$. Setting $U=V$ gives an equality in \eqref{EQ:WTBC_converse_proof_cond1}. For \eqref{EQ:WTBC_converse_proof_cond2}, we have
\begin{align*}
&H_p(Y_1|Z)+I_p(U;Y_2)-I_p(U;Z,Y_1)\\
&=H_p(Y_1|Z)+I_p(V,T;Y_2)-I_p(V,T;Z,Y_1)-I_p(T;Y_2|V)\\
&\quad\quad\quad\quad\quad\quad\quad\quad\quad\quad\quad\quad\quad\quad\quad\quad\quad\mspace{14mu}+I_p(T;Z,Y_1|V)\\
&\geq H_p(Y_1|Z)+I_p(V,T;Y_2)-I_p(V,T;Z,Y_1)\numberthis
\end{align*}
where the last step follows by the implication of the hypothesis \eqref{EQ:WTBC_converse_proof_case11}. Assume that the opposite of \eqref{EQ:WTBC_converse_proof_case11} holds, i.e., 
\begin{equation}
I_p(T;Y_2|V)-I_p(T;Z|V)\geq 0.\label{EQ:WTBC_converse_proof_case2}
\end{equation}
Setting $U=(V,T)$ satisfies \eqref{EQ:WTBC_converse_proof_cond2} with equality, while \eqref{EQ:WTBC_converse_proof_cond2} becomes
\begin{align*}
I_p(U&;Y_2)-I_p(U;Z)\\
&=I_p(V;Y_2)-I_p(V;Z)+I_p(T;Y_2|V)-I_p(T;Z|V)\\
&\geq I_p(V;Y_2)-I_p(V;Z)\numberthis
\end{align*}
where the last inequality holds under hypothesis \eqref{EQ:WTBC_converse_proof_case2}. Concluding, the outer bound described by \eqref{EQ:SL_WTBC_converse_proof_UB} is further contained in $\mathcal{C}^{(\mathsf{SD})}_\mathsf{WT}\left(\mathds{1}_{\{Y_1=y_1(X)\}}p_{Y_2,Z|X}\right)$ from Theorem~\ref{TM:WTBC_capacity}, which established the converse.
\begin{remark}[Theorem 1 from~\cite{Lapidoth_senideterministic2012} for Stochastic Encoders]\label{REM:SD_GPBC_stochastic_encoder}
One important issue to address is that in~\cite{Lapidoth_senideterministic2012}, codes for the GPBC were defined to have a \emph{deterministic encoder}, rather than a stochastic one as considered here. We must, therefore, make sure that this modification does not compromise the validity of the converse proof of Theorem 1 from \cite{Lapidoth_senideterministic2012}. Indeed, taking a closer look at the converse (Section III in~\cite{Lapidoth_senideterministic2012}), we see that the only step that uses the encoder being deterministic is (55), where it is claimed that
\begin{subequations}
\begin{align*}
    \sum_{i=1}^n I\big(M_1,M_2,Z_{[1:i-1]},Z_{[i+1:n]},&Y_{1,[i+1:n]};Y_{1,i}|Z_i\big)\\
    &=\sum_{i=1}^n H\left(Y_{1,i}|Z_i\right)\numberthis\label{EQ:GPBC_lapidoth_deterministic_enc}
\end{align*}
because ``given $(M_1,M_2,Z^n)$, the channel inputs $X^n$ are determined by the encoder, and hence $Y_1^n$ are also determined''. However, since this step is a part of an upper bound (on the sum of rates), we may replace it with
\begin{align*}
    \sum_{i=1}^n I\big(M_1,M_2,Z_{[1:i-1]},Z_{[i+1:n]},&Y_{1,[i+1:n]};Y_{1,i}|Z_i\big)\\
    &\leq\sum_{i=1}^n H\left(Y_{1,i}|Z_i\right),\numberthis\label{EQ:GPBC_lapidoth_deterministic_enc_stochastic}
\end{align*}
which merely uses the non-negativity of conditional entropy and is also valid for stochastic encoders. With that minor modification, one may proceed with the same steps from~\cite[Section III]{Lapidoth_senideterministic2012} to establish the converse when stochastic encoding is allowed.
\end{subequations}
\end{remark}


\subsection{Proof of Theorem~\ref{TM:PD_WTBC_capacity}}\label{SUBSEC:PD_WTBC_capacity_proof}


\subsubsection{\underline{Direct}} Fix a distribution $p_{U,X}\in\mathcal{P}(\mathcal{U}\times\mathcal{X})$ and let $p_{U,X,Y_1,Y_2,Z}\triangleq p_{U,X}p_{Y_1,Y_2,Z|X}$ denote the joint single-letter distribution. We propose a superposition stealth-wiretap code for the PD-WTBC and show that it has an exponentially decaying error probability and effective secrecy metric (see Definition \eqref{DEF:WTBC_achievability_classic}). We then use Part 2 of Proposition~\ref{PROP:Achievability_Relation} to deduce achievability in the sense of Definition~\ref{DEF:WTBC_achievability}.

	
\textbf{Codebook $\bm{\mathcal{B}_n}$:} We use a superposition codebook where $M_1$ and $M_2$ are carried by the outer and inner layers, respectively. Each layer is injected with sufficient randomness that enables to conceal the corresponding message from the eavesdropper.
	
Define the index sets $\mathcal{W}_j^{(n)}\triangleq\big[1:2^{n\tilde{R}_j}\big]$, for $j=1,2$. Let $\mathsf{B}_U^{(n)}\triangleq\big\{\mathbf{U}(m_2,w_2)\big\}_{(m_2,w_2)\in\mathcal{M}_2^{(n)}\times\mathcal{W}_2^{(n)}}$ be a random inner layer codebook, which is a set of random vectors of length $n$ that are i.i.d. according to $p_U^n$. An outcome of $\mathsf{B}_U^{(n)}$ is denoted by $\mathcal{B}_U^{(n)}\triangleq\big\{\mathbf{u}(i)\big\}_{(m_2,w_2)\in\mathcal{M}_2^{(n)}\times\mathcal{W}_2^{(n)}}$. 

To describe the outer layer codebook, fix $\mathcal{B}_U^{(n)}$ and for every $(m_2,w_2)\in\mathcal{M}_2^{(n)}\times\mathcal{W}_2^{(n)}$, let $\mathsf{B}_X^{(n)}(m_2,w_2)\triangleq\big\{\mathbf{V}(m_1,w_1|m_2,w_2)\big\}_{(m_1,w_1)\in\mathcal{M}_1^{(n)}\times\mathcal{W}_1^{(n)}}$ be a collection of i.i.d. random vectors of length $n$ with distribution $q^n_{X|U=\mathbf{u}(m_2,w_2)}$. For each $(m_2,w_2)\in\mathcal{M}_2^{(n)}\times\mathcal{W}_2^{(n)}$, an outcome of $\mathsf{B}_X^{(n)}(m_2,w_2)$ (given an inner layer codebook $\mathcal{B}_U^{(n)}$) is denoted by $\mathcal{B}_X^{(n)}(m_2,w_2)\triangleq\big\{\mathbf{v}(m_1,w_1|m_2,w_2)\big\}_{(m_1,w_1)\in\mathcal{M}_1^{(n)}\times\mathcal{W}_1^{(n)}}$. We also set $\mathsf{B}_X=\big\{\mathsf{B}_X(m_2,w_2)\big\}_{(m_2,w_2)\in\mathcal{M}_2^{(n)}\times\mathcal{W}_w^{(n)}}$ and denote its realizations by $\mathcal{B}_X$. Finally, a random superposition codebook is given by   $\mathsf{B}_n=\Big\{\mathsf{B}_U^{(n)},\mathsf{B}_X^{(n)}\Big\}$, while $\mathcal{B}_n=\Big\{\mathcal{B}_U^{(n)},\mathcal{B}_X^{(n)}\Big\}$ denotes a fixed codebook.
    
Let $\mathfrak{B}_n$ be the set of all possible outcomes of $\mathsf{B}_n$. The above codebook construction induces a distribution $\mu\in\mathcal{P}(\mathfrak{B}_n)$ over the codebook ensemble. For every $\mathcal{B}_n\in\mathfrak{B}_n$, we have
\begin{align*}
    \mu(\mathcal{B}_n)&=\mspace{-20mu} \prod_{(m_2,w_2)\in\mathcal{M}_2^{(n)}\times\mathcal{W}_2^{(n)}}\mspace{-20mu}p^n_U\big(\mathbf{u}(m_2,w_2)\big)\\& \times\mspace{-38mu}\prod_{\substack{\big(\hat{m}_2,\hat{w}_2,m_1,w_1\big)\\\in\mathcal{M}_2^{(n)}\times\mathcal{W}_2^{(n)}\times\mathcal{M}_1^{(n)}\times\mathcal{W}_1^{(n)}}}\mspace{-40mu}p^n_{X|U}\Big(\mathbf{x}\big(m_1,w_1|\hat{m}_2,\hat{w}_2\big)\Big|\mathbf{u}(\hat{m}_2,\hat{w}_2)\Big)\numberthis\label{EQ:codebook_probability}
\end{align*}
The encoder and both decoders are described next for a given superposition codebook $\mathcal{B}_n\in\mathfrak{B}_n$.
	

\textbf{Encoder $\bm{f^{(\mathcal{B}_n)}}$:} Given $(m_1,m_2)\in\mathcal{M}_1^{(n)}\times\mathcal{M}_2^{(n)}$, the encoder chooses $(w_1,w_2)\in\mathcal{M}_1^{(n)}\times\mathcal{M}_2^{(n)}$ uniformly at random and independently of the messages. The channel input sequence is set as $\mathbf{x}(m_1,w_1|m_2,w_2)\in\mathcal{B}_X^{(n)}(m_2,w_2)$. Accordingly, the (stochastic) encoding function $f_n:\mathcal{M}_1^{(n)}\times\mathcal{M}_2^{(n)}\to\mathcal{P}(\mathcal{X}^n)$ is
\begin{align*}
    &f_n^{(\mathcal{B}_n)}(\mathbf{x}|m_1,m_2)\\
    &=\frac{1}{\big|\mathcal{W}_1^{(n)}\big\|\mathcal{W}_2^{(n)}\big|}\sum_{(w_1,w_2)\in\mathcal{W}_1^{(n)}\times\mathcal{W}_2^{(n)}}\mathds{1}_{\big\{\mathbf{x}=\mathbf{x}(m_1,w_1|m_2,w_2)\big\}}.\numberthis\label{EQ:main_proof_encoder}
\end{align*}

	
\textbf{Decoder 1 $\bm{\phi_1^{(\mathcal{B}_n)}}$:} Upon observing $\mathbf{y}_1\in\mathcal{Y}_1^n$,  Decoder 1 searches for a unique tuple $(\hat{m}_1,\hat{m}_2,\hat{w}_1,\hat{w}_2)\in\mathcal{M}_1^{(n)}\times\mathcal{M}_2^{(n)}\times
\mathcal{W}_1^{(n)}\times\mathcal{W}_2^{(n)}$ such that $\Big(\mathbf{u}(\hat{m}_2,\hat{m}_2),\mathbf{x}(\hat{m}_1,\hat{w}_1|\hat{m}_2,\hat{w}_2),\mathbf{y}_1\Big)\in\mathcal{T}_\epsilon^{n}(p_{U,X,Y_1})$. If an appropriate tuple is found, set $\phi_1^{(\mathcal{B}_n)}(\mathbf{y}_1)=\hat{m}_1$; otherwise $\phi_1^{(\mathcal{B}_n)}(\mathbf{y})=1$.
	
\textbf{Decoder 2 $\bm{\phi_2^{(\mathcal{B}_n)}}$:} Upon observing $\mathbf{y}_2\in\mathcal{Y}_2^n$, Decoder 2 searches for a unique pair $(\hat{m}_2,\hat{w}_2)\in\mathcal{M}_2^{(n)}\times\mathcal{W}_2^{(n)}$ such that $
    \Big(\mathbf{u}(\hat{m}_2,\hat{m}_2),\mathbf{y}_2\Big)\in\mathcal{T}_\epsilon^{n}(p_{U,Y_2})$. The output of the decoding function is defined as $\phi_2^{(\mathcal{B}_n)}(\mathbf{y}_2)=\hat{m}_2$, if a unique pair is found, and as $\phi_2^{(\mathcal{B}_n)}(\mathbf{y}_2)=1$, otherwise.
	
The triple $(f^{(\mathcal{B}_n)},\phi_1^{(\mathcal{B}_n)},\phi_2^{(\mathcal{B}_n)})$ defined with respect to any codebook $\mathcal{B}_n\in\mathfrak{B}_n$ constitutes an $(n,R_1,R_2)$-code $c_n$ for the WTBC. For any codebook $\mathcal{B}_n\in\mathfrak{B}_n$, the induced joint distribution is
\begin{align*}
    &P^{(\mathcal{B}_n)}(m_1,m_2,w_1,w_2,\mathbf{u},\mathbf{x},\mathbf{y}_1,\mathbf{y}_2,\mathbf{z},\hat{m}_1,\hat{m}_2)\\
    &= \frac{1}{\prod\limits
    _{j=1,2}\big|\mathcal{M}_j^{(n)}\big|\big|\mathcal{W}_j^{(n)}\big|}\mathds{1}_{\big\{\mathbf{u}=\mathbf{u}(m_2,w_2)\big\}\cap \big\{\mathbf{x}=\mathbf{x}(m_1,w_1|m_2,w_2)\big\}}\\
    & \mspace{50mu}\times p^n_{Y_1,Y_2,Z|X}(\mathbf{y}_1,\mathbf{y}_2,\mathbf{z}|\mathbf{x})\mathds{1}_{\bigcap\limits_{j=1,2}\big\{\hat{m}_j=\phi_j^{(\mathcal{B}_n)}(\mathbf{y}_j)\big\}}.\numberthis\label{EQ:main_proof_induced_PMF}
\end{align*}
    
Set $\mathsf{P}_\mathsf{e}(\mathcal{B}_n)\triangleq \mathbb{P}_{P^{(\mathcal{B}_n)}}\big((\hat{M}_1,\hat{M}_2)\neq(M_1,M_2)\big)$ as the error probability of a code $c_n$ (induced by $\mathcal{B}_n$).

    
    
\textbf{Expected Probability of Error Analysis:} The above described codebook, encoder and decoders correspond to the classic superposition coding scheme for BCs with $\tilde{M}_j\triangleq(M_j,W_j)$, for $j=1,2$, in the role of the $j$-th message. Via standard joint-typicality analysis of the error probability (see, e.g,~\cite[Section 5.3.1]{ElGamal2011}), there exists a $\tau_1>0$ such that $\mathbb{E}_\mu\mathsf{P}_\mathsf{e}(\mathsf{B}_n)\leq e^{-n\tau_1}$, provided that $(R_1,R_2)$ satisfy
\begin{subequations}
\begin{align}
    R_1+\tilde{R}_1&<I_p(X;Y_1,Z|U)\label{EQ:PD_achievable_proof_RB1}\\
    R_2+\tilde{R}_2&<I_p(U;Y_2)\label{EQ:PD_achievable_proof_RB2}\\
    R_1+\tilde{R}_1+R_2+\tilde{R}_2&<I_p(U,X;Y_1,Z).\label{EQ:PD_achievable_proof_RB12}
\end{align}\label{EQ:PD_achievable_proof_RB}%
\end{subequations}
Note that the sum of \eqref{EQ:PD_achievable_proof_RB1} and \eqref{EQ:PD_achievable_proof_RB2} gives $R_1+\tilde{R}_1+R_2+\tilde{R}_2< I_p(U;Y_2) + I_p(X;Y_1,Z|U)$; this makes \eqref{EQ:PD_achievable_proof_RB12} redundant. The redundancy is due to the assumption that the PD-WTBC satisfies $Y_2- Y_1- X$, which, in turn, implies $I_p(U;Y_1,Z)\geq I_p(U;Y_1)\geq I_p(U;Y_2)$.


\textbf{Effective Secrecy Analysis:} We analyze the expected value of the effective secrecy metric. Let $p_Z$ be the $Z$-marginal of the single-letter distribution $p_{U,X,Y_1,Y_2,Z}$ and observe that
\begin{align*}
    \mathbb{E}_\mu \mathsf{D}\Big(P^{(\mathsf{B}_n)}_{M_1,M_2,\mathbf{Z}}&\Big\|p^{(\mathsf{U})}_{\mathcal{M}_1^{(n)}\times\mathcal{M}_2^{(n)}}p_Z^n\Big)\\
    &\stackrel{(a)}=\mathbb{E}_\mu \mathsf{D}\Big(P^{(\mathsf{B}_n)}_{\mathbf{Z}|M_1,M_2}\Big\|p_Z^n\Big|p^{(\mathsf{U})}_{\mathcal{M}_1^{(n)}\times\mathcal{M}_2^{(n)}}\Big)\\
    &\stackrel{(b)}=\mathbb{E}_\mu \mathsf{D}\Big(P^{(\mathsf{B}_n)}_{\mathbf{Z}|M_1=1,M_2=1}\Big\|p_Z^n\Big)\numberthis\label{EQ:effective_sec_analysis}
\end{align*}
where (a) is because for every $\mathcal{B}_m\in\mathfrak{B}_n$ we have $P^{(\mathsf{B}_n)}_{M_1,M_2}=p^{(\mathsf{U})}_{\mathcal{M}_1^{(n)}\times\mathcal{M}_2^{(n)}}$ and the KL divergence chain rule, while (b) uses symmetry (with respect to $(m_1,m_2)$). 

The KL divergence term on the RHS of \eqref{EQ:effective_sec_analysis} falls under the framework of the strong soft-covering lemma for superposition codes from~\cite[Lemma 1]{Goldfeld_SDWTC2016}, which implies that if \begin{subequations}
\begin{align}
    \tilde{R}_1&>I_p(X;Z|U)\label{EQ:PD_achievable_proof_covering_RB1}\\
    \tilde{R}_2&>I_p(U;Z).\label{EQ:PD_achievable_proof_covering_RB2}
\end{align}\label{EQ:PD_achievable_proof_covering_RB}%
\end{subequations}
then there exists $\tau_2>0$ such that
$\mathbb{E}_\mu \mathsf{D}\Big(P^{(\mathsf{B}_n)}_{\mathbf{Z}|M_1=1,M_2=1}\Big\|p_Z^n\Big)\leq e^{-n\tau_2}$.


\textbf{Finalization:} We have that if $(R_1,R_2)$ and $p_{U,X,Y_1,Y_2,Z}$ are such that \eqref{EQ:PD_achievable_proof_RB1}-\eqref{EQ:PD_achievable_proof_RB2} and \eqref{EQ:PD_achievable_proof_covering_RB} are satisfied, then 
\begin{subequations}
\begin{align}
    \mathbb{E}_\mu\mathsf{P}_\mathsf{e}(\mathsf{B}_n)&\leq e^{-n\tau_1}\\
    \mathbb{E}_\mu \mathsf{D}\Big(P^{(\mathsf{B}_n)}_{M_1,M_2,\mathbf{Z}}\Big\|p^{(\mathsf{U})}_{\mathcal{M}_1^{(n)}\times\mathcal{M}_2^{(n)}}p_Z^n\Big)&\leq e^{-n\tau_2},
\end{align}
\end{subequations}
for all sufficiently large $n$. Through standard existence arguments (e.g., via Markov's inequality), we conclude that there exists a sequence of superposition codebooks $\{\mathcal{B}_n\}_{n\in\mathbb{N}}$ (each giving rise to an $(n,R_1,R_2)$-code $c_n$ with induced distribution $P^{(c_n)}$) satisfying \eqref{EQ:WTBC_cachievability_exponential} from Proposition~\ref{PROP:Achievability_Relation}. Invoking Part 2 of Proposition \ref{PROP:Achievability_Relation}, we conclude that 
\begin{align*}
    \left\|P^{(c_n)}_{M_1,M_2,\hat{M}_1,\hat{M}_2,\mathbf{Z}}-p^{(\mathsf{U})}_{\mathcal{M}^{(n)}_1\times\mathcal{M}^{(n)}_2}\mathds{1}_{\big\{(\hat{M}_1,\hat{M}_2)=(M_1,M_2)\big\}}q_Z^n\right\|_\mathsf{TV}&\\&\mspace{-100mu}\leq e^{-n\gamma},\numberthis
\end{align*}
for some $\gamma>0$ and all sufficiently large $n$. Applying Fourier-Motzkin Elimination on \eqref{EQ:PD_achievable_proof_RB1}-\eqref{EQ:PD_achievable_proof_RB2} and \eqref{EQ:PD_achievable_proof_covering_RB} to eliminate $\tilde{R}_1$ and $\tilde{R}_2$ establishes the direct part of Theorem~\ref{TM:PD_WTBC_capacity}.


\subsubsection{\underline{Converse}}\label{SUBSUBSEC:PD_WTBC_converse_proof}

The converse proof is similar to that of Theorem~\ref{TM:WTBC_capacity} (for the SD-WTBC) given in Section~\ref{SUBSUBSEC:WTBC_converse_proof}. Namely, we use the converse proof of the coding theorem for the analogous PD-GPBC with an informed receiver~\cite[Theorem 3]{Steinberg_Degraded_BC_State2005} and invoke Proposition~\ref{PROP:WTGPBC_codes_relations} to relate it to the considered PD-WTBC. To avoid repetitiveness, we only outline that main steps of the derivation. 

For an achievable rate pair $(R_1,R_2)\in\mathbb{R}_+^2$, let $\{c_n\}_{n\in\mathbb{N}}$ be a sequence of $(n,R_1,R_2)$-codes satisfying \eqref{EQ:WTBC_achievability_def}. Proposition~\ref{PROP:WTGPBC_codes_relations} produces a sequence of $(n,R_1,R_2)$-codes $\{b_n\}_{n\in\mathbb{N}}$ for the 
$\big(\mathcal{X},\mathcal{Y}_1,\mathcal{Y}_2,\mathcal{Z},p_{Y_1,Z|X}p_{Y_2|Y_1}\big)$ PD-WTBC with an informed receiver, each inducing a joint distribution $Q^{(b_n)}$ (see \eqref{EQ:GPBC_induced_PMF}) such that: (1)  $\big\|P^{(c_n)}-Q^{(b_n)}\big\|_\mathsf{TV}\leq e^{-n\gamma}$, for any sufficiently large $n$; and (2) $\mathsf{P}_\mathsf{e}(b_n)\xrightarrow[n\to\infty]{}0$.

The converse proof of Theorem 3 from~\cite{Steinberg_Degraded_BC_State2005} is given in Section V-D therein and shows that
\begin{subequations}
\begin{align}
    R_1&\leq \frac{1}{n}\sum_{i=1}^nI_{Q^{(b_n)}}(X_i;Y_{1,i}|M_2,Y_{2,[1:i-1]},Z_{[i+1:n]},Z_i)\mspace{-3mu}+\mspace{-3mu}\epsilon_n\label{EQ:PD_WTBC_converse_proof_UB1}\\
    R_2&\leq \frac{1}{n}\sum_{i=1}^n\Big[I_{Q^{(b_n)}}\big(M_2,Y_{2,[1:i-1]},Z_{[i+1:n]};Y_{2,i}\big)\nonumber\\
    &\quad\quad\mspace{6mu}-I_{Q^{(b_n)}}\big(M_2,Y_{2,[1:i-1]},Z_{[i+1:n]};Z_i\big)\Big]+\epsilon_n\label{EQ:PD_WTBC_converse_proof_UB2}
\end{align}\label{EQ:PD_WTBC_converse_proof_UB}%
\end{subequations}
where $\epsilon_n\triangleq\frac{2}{n}+\max\{R_1,R_2\}\cdot\mathsf{P}_\mathsf{e}(b_n)\xrightarrow[n\to\infty]{}0$. Since $P^{(c_n)}$ and $Q^{(b_n)}$ are exponentially close in total variation, Lemma~\ref{LEMMA:MI_continuity} further implies that for some $\delta>0$ and any $n$ large enough
\begin{subequations}
\begin{align}
    R_1&\leq \frac{1}{n}\sum_{i=1}^nI_{P^{(c_n)}}(X_i;Y_{1,i}|U_i,Z_i)+\epsilon_n+e^{-n\delta}\label{EQ:P_PD_WTBC_converse_proof_UB1}\\
    R_2&\leq \frac{1}{n}\sum_{i=1}^n\mspace{-2mu}\Big[I_{P^{(c_n)}}(U_i;Y_{2,i})-I_{P^{(c_n)}}(U_i;Z_i)\Big]\mspace{-3mu}+\mspace{-3mu}\epsilon_n\mspace{-3mu}+\mspace{-3mu}2e^{-n\delta},\label{EQ:P_PD_WTBC_converse_proof_UB2}
\end{align}\label{EQ:P_PD_WTBC_converse_proof_UB}%
\end{subequations}
where $U_i\triangleq \big(M_2,Y_{2,[1:i-1]},Z_{[i+1:n]}\big)_{P^{(c_n)}}$, for $i\in[1:n]$. The structure of $P^{(c_n)}$ in \eqref{EQ:WTBC_induced_PMF} reveals that $(Y_{1,i},Y_{2,i},Z_i) - X_i - U_i$ forms a Markov chain for every $i\in[1:n]$. The PD assumption further gives that the chain $Y_{2,i} - Y_{1,i} - (Z_i,X_i,U_i)$ is Markov. Letting $n$ tend to infinity in \eqref{EQ:P_PD_WTBC_converse_proof_UB} established the converse of Theorem~\ref{TM:PD_WTBC_capacity}.

\begin{remark}[Theorem 3 from~\cite{Steinberg_Degraded_BC_State2005} for Stochastic Encoders]
In accordance to the discussion from Remark~\ref{REM:SD_GPBC_stochastic_encoder}, for an analogy based converse proof of Theorem~\ref{TM:PD_WTBC_capacity} to work, one has to verify that the result of~\cite[Theorem 3]{Steinberg_Degraded_BC_State2005} applies to stochastic encoders. Examining the steps of the proof therein (see~\cite[Section V-D]{Steinberg_Degraded_BC_State2005}), it is readily seen that the derivation holds true regardless of whether the encoder is deterministic or stochastic.
\end{remark}


\section{Concluding Remarks and Future Work}\label{SEC:summary}

This paper formalized an analogy that connects wiretap and GP channel coding problems. 
We showed that given a wiretap model and a good sequence of stealth-wiretap codes, one can construct an analogous GPC whose capacity upper/outer bounds that of the WTC. 
Consequently, any upper/outer bound on the GPC capacity also upper/outer bounds the secrecy-capacity of the original WTC. This framework extends to broadcast channels and was used to derive new WTBC capacity results. Directly extending the similarity between the capacity expressions of the classic wiretap and GPCs (see \eqref{EQ:WTC_capacity_intro} and \eqref{EQ:GPC_capacity_intro}), our obtained admissible WTBC regions correspond to those of their GP analogs. The regions of two analogous setups are described by the same information-theoretic bounds on the transmission rates, but the underlying distributions are different (albeit related). The derivations of the WTBC secrecy-capacity results demonstrate how to use the analogy to prove converses. We postulate that additional results can be derived based on the analogy. Promising directions include 3-receiver WTBCs (based on GP results such as the one from \cite{ramachandran2017capacity}) and possibly a new WTC strong converse proof based on the strong converse for the GPC \cite{hayashi2014strong} (a strong converse is known only for the \emph{degraded} WTC \cite{tyagi2009gelfand}, and was derived via an active hypothesis testing technique). We leave these explorations for future work.

Other appealing future research directions include extensions of the analogy framework to additional multiuser scenarios, such as MACs. The cognitive GP-MAC was solved in~\cite{GP_MAC_Shamai2008}, while the cognitive wiretap MAC (WT-MAC), studied in~\cite{WT_MAC_Simeone2008}, remains an open problem. Extending the analogy to encompass MACs could help resolving the question of the cognitive WT-MAC's secrecy-capacity region. Developing the MAC analogy requires addressing several delicate issues such as the correlation between the encoders of the cognitive WT-MAC that is not present in the GP version of the problem. It would also be interesting to explore an analogy framework that accounts for the GP-WTC, i.e., a state-dependent WTC with non-causal encoder CSI. This challenging problem received considerable attention in recent years~\cite{Mitrpant_Gaussian_SDWTC2006,SDWTC_Chen_HanVinck2006,SDWTC_2Sided_Liu2007,Goldfeld_SDWTC2016,El-Halabi_SDWTC2012,Bafghi_SDWTC2012}. While various achievability claims have been derived, and although the secrecy-capacity is known in some special cases, good upper bounds are scarcely available. One would expect that the analogous GPC would be driven by a pair of states: one that corresponds to the GP-WTC's state and the other to the eavesdropper's signal. Although the exact details have yet been fully understood, we note that state-dependent channels driven by a pair of states have several known solutions in the literature~\cite{cover_duality2002,khina_GPC2State}. An analogy framework between these two setups would open the door to deriving improved (possibly tight) upper bounds of the GP-WTC's secrecy-capacity. Finally, since this paper only establishes a uni-directional relation between the wiretap and GPCs, a natural future research trajectory is to explore if and when an opposite relation holds true.

\section*{Acknowledgements}
The authors thank the Associate Editor and the anonymous reviewers for the thorough review process that helped us greatly improve the presentation of this paper. 


\appendices

\bibliographystyle{unsrt}
\bibliographystyle{IEEEtran}
\bibliography{ref}

\begin{IEEEbiographynophoto}{Ziv Goldfeld}

Dr. Ziv Goldfeld is currently a postdoctoral fellow at the Laboratory for
Information and Decision Systems (LIDS) at MIT. He graduated with a B.Sc.
summa cum laude, an M.Sc. summa cum laude, and a Ph.D. in Electrical and
Computer Engineering from Ben-Gurion University, Israel, in 2012, 2015 and
2018, respectively. His research interest include theoretical machine
learning, information theory, complex dynamical systems, high-dimensional
and nonparametric statistics and applied probability. Honors include the

(S'13-M'17) received his B.Sc.\@ (summa cum laude), M.Sc.\@ (summa cum laude) and Ph.D. degrees in Electrical and Computer Engineering from the Ben-Gurion University, Israel, in 2012, 2014 and 2017,  respectively. He is currently a postdoctoral fellow at the Laboratory for Information and Decision Systems (LIDS) at MIT.

Ziv is a recipient of several awards, among them the are Rothschild postdoctoral fellowship, the Ben Gurion University postdoctoral fellowship, the Feder Award, a best student paper award in the IEEE 28-th Convention of Electrical and Electronics Engineers in Israel, the Basor fellowship for outstanding students in the direct Ph.D. program, the Lev-Zion fellowship and the Minerva Short-Term Research Grant (MRG).
\end{IEEEbiographynophoto}

\begin{IEEEbiographynophoto}{Haim H. Permuter}
(M'08-SM'13) received his B.Sc.\@ (summa cum laude) and M.Sc.\@ (summa cum laude) degrees in Electrical and Computer Engineering from the Ben-Gurion University, Israel, in 1997 and 2003, respectively, and the Ph.D. degree in Electrical  Engineering from Stanford University, California in 2008.

Between 1997 and 2004, he was an officer at a research and development unit of the Israeli Defense Forces. Since 2009 he is with the department of Electrical and Computer Engineering at Ben-Gurion University where he is currently  a professor, Luck-Hille Chair in Electrical Engineering. Haim also serves as head of the communication track in his department.

Prof. Permuter is a recipient of several awards, among them the
Fullbright Fellowship, the Stanford Graduate Fellowship (SGF), Allon
Fellowship, and and the U.S.-Israel Binational Science Foundation
Bergmann Memorial Award. Haim served on the editorial boards of the IEEE Transactions on Information Theory in 2013-2016.

\end{IEEEbiographynophoto}\end{document}